\documentclass[envcountsect,orivec]{llncs} 
\usepackage{etex} 
\usepackage[]{graphicx}
\newif\ifignore 
\ignorefalse
\newcommand{\auxproof}[1]{
  \ifignore\mbox{}\newline
  \textbf{BEGIN: AUX-PROOF} \dotfill\newline
  {#1}\mbox{}\newline
  \textbf{END: AUX-PROOF}\dotfill\newline
  \fi}

\usepackage{times}

\usepackage{theorem}
\usepackage{fancybox,amssymb,amstext,amsmath,stmaryrd,wasysym,cite,proof,mathtools,multirow,stackrel}
\SetSymbolFont{stmry}{bold}{U}{stmry}{m}{n}
\SetSymbolFont{wasy}{bold}{U}{wasy}{m}{n}
\usepackage{algorithm}
\usepackage{algpseudocode}
\makeatletter
\algrenewcommand\ALG@beginalgorithmic{\footnotesize}
\makeatother
\usepackage[pdftex,all]{xy}
\usepackage{xspace}
\allowdisplaybreaks[1] 

\xyoption{v2}
\xyoption{curve}
\xyoption{2cell}
\SelectTips{cm}{}  
\UseAllTwocells
\SilentMatrices
\def\labelstyle{\scriptstyle}
\def\twocellstyle{\scriptstyle}
\newdir{ >}{{}*!/-8pt/@{>}}  
\newdir{|>}{%
  !/1.6pt/@{|}*:(1,-.2)@^{>}*:(1,+.2)@_{>}}
\newdir{pb}{:(1,-1)@^{|-}}
\def\pb#1{\save[]+<20 pt,0 pt>:a(#1)\ar@{pb{}}[]\restore}

\usepackage{wrapfig}
\theorembodyfont{\itshape}
\newtheorem{mytheorem}{Theorem}[section]
\newtheorem{mylemma}[mytheorem]{Lemma}
\newtheorem{myproposition}[mytheorem]{Proposition}

\theorembodyfont{\rmfamily}

\newtheorem{myremark}[mytheorem]{Remark}

\newtheorem{mydefinition}[mytheorem]{Definition}

\spnewtheorem*{myproof}{Proof}{\itshape}{\rmfamily}

\def\myqed{\qed}

\newcommand{\R}{{\mathbb{R}}}

\newcommand{\Var}{\mathbf{Var}}
\newcommand{\ttrue}{\mathrm{t{\kern-1.5pt}t}}
\newcommand{\ffalse}{\mathrm{f{\kern-1.5pt}f}}

\newcommand{\place}{\underline{\phantom{n}}\,}

\newcommand{\STL}{\textbf{STL}}
\newcommand{\Fml}{\mathbf{Fml}}
\newcommand{\FmlNNF}{\mathbf{Fml}_{\mathrm{NNF}}}
\newcommand{\AP}{\mathbf{AP}}
\newcommand{\UntilOp}[1]{\mathbin{\mathcal{U}_{#1}}}
\newcommand{\Release}[1]{\mathbin{\mathcal{R}_{#1}}}
\newcommand{\TUntil}[1]{\mathbin{\overline{\mathcal{U}}_{#1}}}
\newcommand{\TRelease}[1]{\mathbin{\overline{\mathcal{R}}_{#1}}}
\newcommand{\DiaOp}[1]{\Diamond_{#1}}
\newcommand{\BoxOp}[1]{\square_{#1}}
\newcommand{\TDiaOp}[1]{\overline{\Diamond}_{#1}}
\newcommand{\TBoxOp}[1]{\overline{\square}_{#1}}
\newcommand{\Robust}[2]{{ \llbracket #1,\, #2 \rrbracket}}
\newcommand{\sem}[1]{\llbracket #1 \rrbracket} 
\newcommand{\Defeq}{\triangleq}

\newcommand{\Lim}[1]{{\displaystyle{\lim_{#1}}}}
\newcommand{\Vee}[1]{{{\bigsqcup_{#1}}}}
\newcommand{\Wedge}[1]{{{\bigsqcap_{#1}}}}
\newcommand{\Frac}[2]{{\displaystyle{\frac{#1}{#2}}}}
\newcommand{\Int}{\displaystyle \int}

\newcommand{\AvSTL}{\textbf{AvSTL}}
\newcommand{\Max}{\mathsf{max}}
\newcommand{\Min}{\mathsf{min}}

\newcommand{\C}{\mathcal{C}}

\newcommand{\Rnn}{\R_{\ge 0}}
\newcommand{\Rnp}{\R_{\le 0}}
\newcommand{\Succ}{Succ.}

 \title{
Time Robustness in MTL and
\\Expressivity in
  Hybrid System Falsification
}
 \author{
Takumi Akazaki
 \inst{1,2}
  \and
Ichiro Hasuo
 \inst{1}
 }
 \institute{
    The University of Tokyo, Japan
    \and
    JSPS Research Fellow
}

\begin{document}

\maketitle

\begin{abstract}
  Building on the work by Fainekos and Pappas and the one by Donz\'{e}
  and Maler, we introduce $\AvSTL$, an extension of metric interval
  temporal logic by \emph{averaged} temporal operators.  Its
  expressivity in capturing both space and time robustness helps solving
  \emph{falsification} problems (searching for a critical path in
  hybrid system models); it does so by communicating a designer's intention more
  faithfully to the stochastic optimization engine employed in a
  falsification solver.  We also introduce a sliding window-like
  algorithm that keeps the cost of computing truth/robustness values tractable.

\end{abstract}

\section{Introduction}\label{sec:introduction}
\paragraph{Model-Based Development of Hybrid Systems}
The demand for quality assurance of \emph{cyber-physical systems (CPS)} is
ever-rising, now that computer-controlled artifacts---cars, aircrafts,
and so on---serve diverse safety-critical tasks 
everywhere in our daily lives.  In the industry practice of CPS design, deployment
of \emph{model-based development (MBD)} has become a norm. 
In MBD, (physical and costly) testing workbenches are replaced 
by (virtual and cheap) \emph{mathematical models}; 
and this reduces by a great deal the cost of running 
a \emph{development cycle}---design, implementation, evaluation, and redesign.

One of the distinctive features of CPS is that they are \emph{hybrid systems} 
and combine discrete and continuous
dynamics. For MBD of such systems the software \emph{Simulink} has emerged as an
industry standard. In Simulink a designer models a system using block
diagrams---a formalism  strongly influenced by \emph{control theory}---and 
runs \emph{simulation}, that is, numerical solution of the system's dynamics.

\paragraph{Falsification}
 The models of most real-world hybrid systems are believed to be beyond
the reach of formal verification.  While this is certainly the case with 
systems as big as a whole car,
a single component of it (like automatic transmission or an engine
controller)  overwhelms the scalability of
the state-of-art formal verification techniques, too.

What is worse, hybrid
system models tend to have \emph{black-box components}. An example is
fuel combustion in an engine.  Such chemical reactions are not easy to
model with
ODEs, and are therefore commonly represented in a Simulink model 
by a \emph{look-up table}---a big table of values obtained by  physical measurements~\cite{DBLP:conf/hybrid/JinDKUB14,HoxhaAF14arch1}. 
The lack of structure in a look-up table poses a
challenge to formal verification:  each entry of the table
calls for separate treatment; and this 
easily leads to state-space explosion.

Under such circumstances,  \emph{falsification} by
stochastic optimization has proved to be a viable
approach to quality
assurance~\cite{DBLP:conf/tacas/AnnpureddyLFS11,DBLP:conf/hybrid/JinDKUB14,HoxhaAF14arch1}. The
problem is formulated as follows:
\begin{quote}
  \underline{\bfseries The falsification problem}

  \begin{tabular}{ll}
    \textbf{Given:} 
    & a \emph{model} $\mathcal{M}$ (a function from an input signal \\
    & to  an output signal), and\\
    & a \emph{specification} $\varphi$ (a temporal formula),\\
    \textbf{Answer:} 
    & a \emph{critical path}, that is, an input signal $\sigma_{\mathrm{in}}$ such\\
    & that the output $\mathcal{M}(\sigma_{\mathrm{in}})$ does not satisfy $\varphi$ 
  \end{tabular}
  \includegraphics[clip,trim=0cm 23.5cm 24.8cm 0cm,width=.24\textwidth]{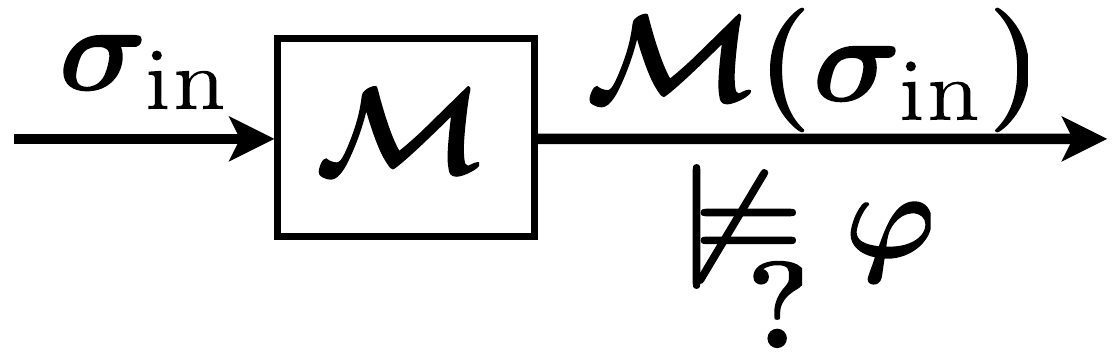}
\end{quote}
Unlike \emph{testing} or \emph{monitoring}---where input $\sigma_{\mathrm{in}}$ is given and
we check if $\mathcal{M}(\sigma_{\mathrm{in}})\models \varphi$---a
falsification solver employs stochastic optimization techniques (like
the Monte-Carlo ones) and iteratively searches for a falsifying input
signal
$\sigma_{\mathrm{in}}$.

Falsification is a versatile tool in MBD of hybrid systems.  It is
capable of searching for counterexamples, hence revealing potential
faults in the design. One can also take, as a specification $\varphi$,
the negation $\lnot \psi$ of a desirable property $\psi$; then
successful falsification amounts to \emph{synthesis} of an input signal that
satisfies $\psi$.  Stochastic optimization used in falsification
typically does not rely on the internal structure of models, therefore the
methodology is suited for models with black-box
components. Falsification is fairly scalable, making it 
a realistic option in
the industrial MBD scenarios; see e.g.~\cite{HoxhaAF14arch1,DBLP:conf/hybrid/JinDKUB14}.


The current work aims at enhancing falsification solvers, notable among which are
S-TaLiRo~\cite{DBLP:conf/tacas/AnnpureddyLFS11}
and
BREACH~\cite{DBLP:conf/cav/Donze10}. 
An obvious way to do so is via improvement of stochastic optimization;
see e.g.~\cite{Sankaranarayanan:2012:FTP:2185632.2185653,Zutshi:2014:MSC:2656045.2656061}. Here we take a different, logical approach.

\paragraph{Robustness in Metric Temporal Logics}
Let us turn to a formalism in which a specification
$\varphi$ is expressed. 
\emph{Metric interval temporal logic} 
($\textbf{MITL}$)~\cite{DBLP:dblp_journals/jacm/AlurFH96}, 
and its adaptation
\emph{signal temporal logic} 
($\textbf{STL}$)~\cite{DBLP:conf/formats/MalerN04}, 
are standard temporal logics for (continuous-time) signals. 
However their conventional
semantics---where satisfaction is Boolean---is not suited for
falsification by stochastic optimization. 
This is because a formula $\varphi$, no matter if it is 
\emph{robustly} satisfied and \emph{barely} satisfied, yields the same
truth value (``true''), making it  not amenable to hill climb-style
optimization.

It is  the introduction of \emph{robust semantics} of
$\textbf{MITL}$~\cite{DBLP:journals/tcs/FainekosP09} that set off the idea of falsification by
optimization. In robust semantics,  a signal $\sigma$ and a
formula $\varphi$ are assigned a continuous truth value $\Robust{\sigma}{\varphi}\in\R$
that designates how robustly the formula is satisfied. Such ``robustness
values'' constitute a sound basis for stochastic optimization.

\begin{wrapfigure}[10]{r}{0pt}
  \centering
  \begin{tabular}{c}
    \includegraphics[width=.3\textwidth,natwidth=640,
    natheight=384]{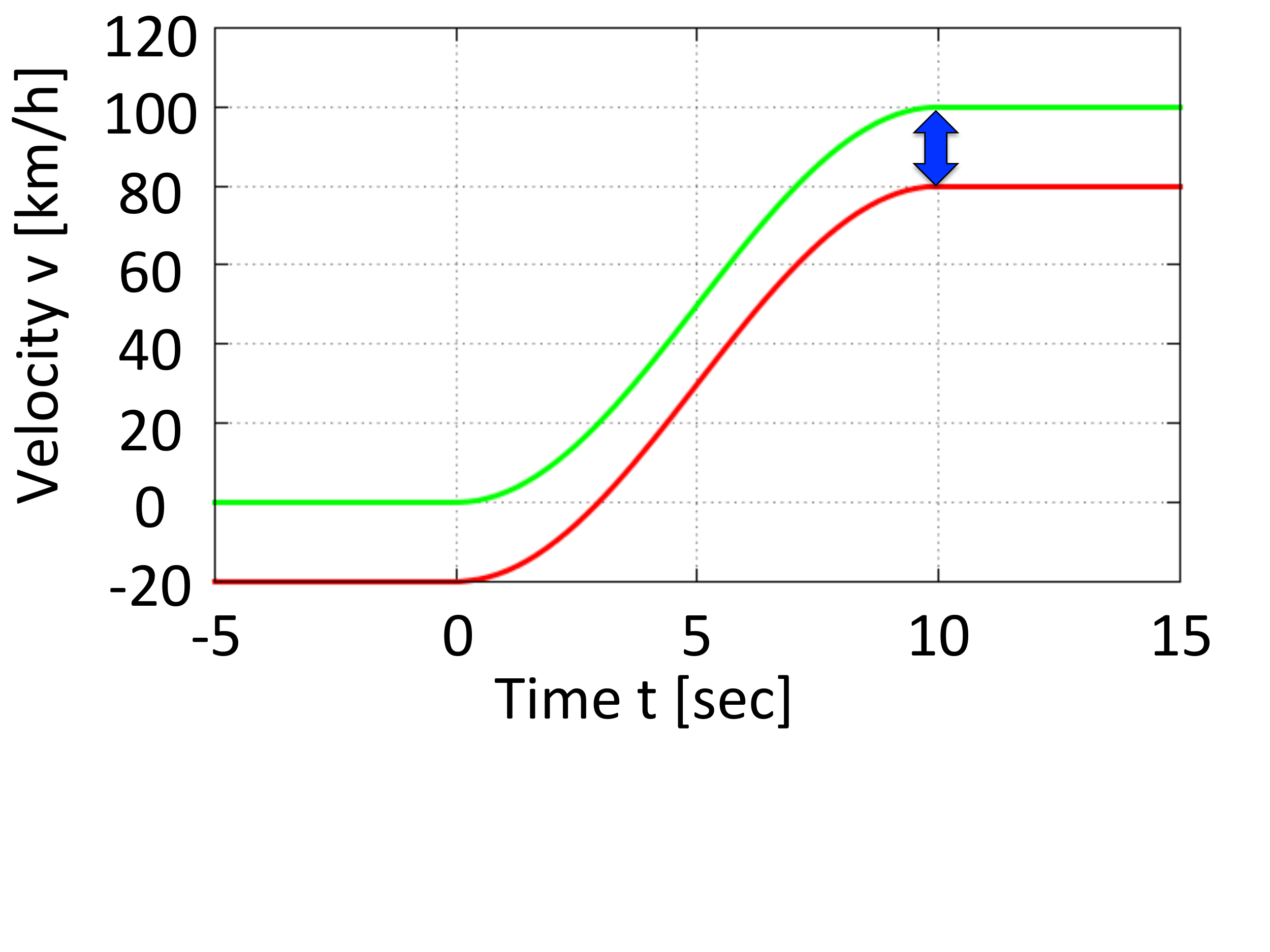}
    \\[-2.5em]
    \includegraphics[width=.3\textwidth,natwidth=640,
    natheight=384]{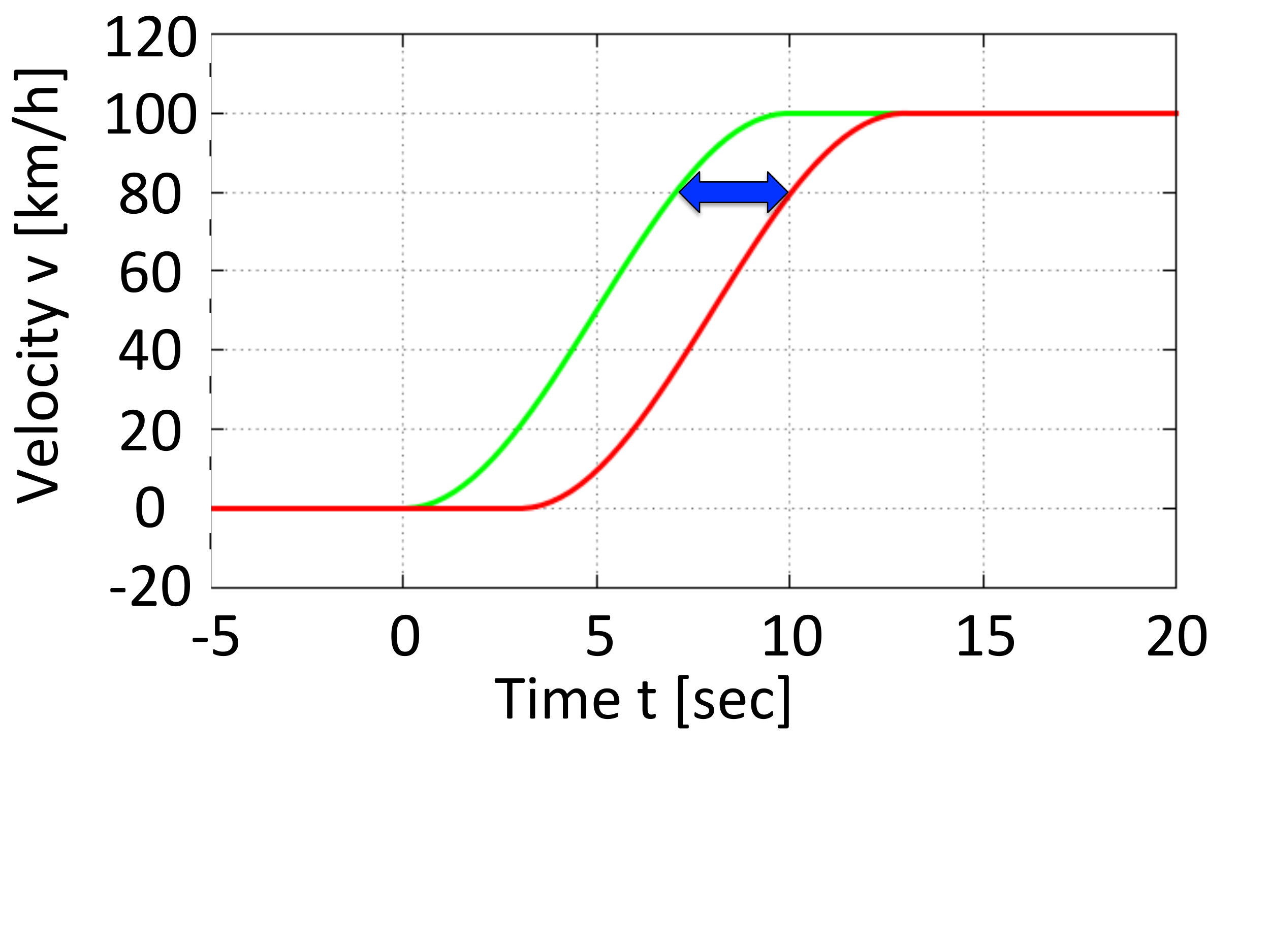}
  \end{tabular}
\end{wrapfigure}
The original
robust semantics in~\cite{DBLP:journals/tcs/FainekosP09} is concerned with
\emph{space} robustness: for example, the truth values of 
$\DiaOp{[0, 10]}(v \geq 80)$ (``the velocity reaches 80 km/h within 10 sec.'')
are $20$ and $0$, for the green and red signals on the right. Therefore
space
robustness is a ``vertical margin'' between a signal and a
specification. An efficient algorithm is proposed in~\cite{DBLP:conf/cav/DonzeFM13} for computing this
notion of robustness.

The notion of robustness is extended
in~\cite{DBLP:conf/formats/DonzeM10} to take  \emph{time} robustness also
into account. Consider the same specification $\DiaOp{[0, 10]}(v \geq
80)$ against the green and red signals on the right. The green one is
more robust since it reaches 80 km/h much earlier than the deadline (10
sec.), while the red one barely makes the deadline.

The current work continues this line of work, with the slogan that 
\emph{expressivity of temporal logic should help falsification}. 
With more expressivity, a designer's concerns that were
previously ignored (much like time robustness was ignored
in~\cite{DBLP:journals/tcs/FainekosP09}) come to be reflected in the
continuous truth value. The latter will in turn help stochastic
optimization by giving additional ``hints.'' We however are in a
\emph{trade-off} situation: the more expressive a logic is, the more expensive
computation of truth values is in general.

\paragraph{Contributions}  We aim at:  a good
balance 
in the last trade-off  between expressivity and computational cost; and thereby
enhancing  falsification solvers by giving 
more ``hints'' to stochastic optimization procedures. Our technical
contributions are threefold.

\textbf{The logic $\AvSTL$.} We introduce \emph{averaged STL} ($\AvSTL$); it is an
extension of $\STL$~\cite{DBLP:conf/formats/MalerN04} by
so-called \emph{averaged temporal operators} like $\TUntil{I}$ and
$\TDiaOp{I}$. The (continuous) truth values of the new operators are defined by  
the average of truth values in a suitable interval. 
We show that this simple extension 
of $\STL$ successfully combines
space and time robustness
in~\cite{DBLP:journals/tcs/FainekosP09,DBLP:conf/formats/DonzeM10}; and 
that its  expressivity covers many common specifications (expeditiousness,
persistence, deadline, etc.) 
encountered in the context of CPS. 

\textbf{An algorithm for computing $\AvSTL$ robustness.} It is natural
to expect that nonlocal temporal operators---like $\UntilOp{I}$, $\DiaOp{I}$ and their 
averaged variants---incur a big performance penalty in  computing
truth values.
For $\STL$ (without averaged modalities) an efficient algorithm is
proposed in~\cite{DBLP:conf/cav/DonzeFM13}; it employs the idea of
the \emph{sliding window minimum algorithm}~\cite{DBLP:journals/njc/Lemire06} 
and achieves complexity
that is linear with respect to the size of an input signal (measured by
the number of timestamps). 

We show that, under mild and realistic
assumptions, the same idea as in~\cite{DBLP:conf/cav/DonzeFM13} can be successfully
employed
to compute $\AvSTL$ truth values with linear complexity.



\textbf{Enhancing S-TaLiRo: implementation and experiments.} 
We use S-TaLiRo and demonstrate that our logic $\AvSTL$ indeed achieves a reasonable balance
between expressivity and computational cost.  We present our
prototype implementation: it takes S-TaLiRo and lets the above algorithm
(called the \emph{$\AvSTL$ evaluator})  replace TaLiRo,
S-TaLiRo's
original engine for computing $\STL$ truth values  (see
Fig.~\ref{fig:staliro} in~\S{}\ref{sec:experiments}).

For its evaluation, 
we pick some 
benchmark models $\mathcal{M}$ and $\STL$ specifications $\varphi$---they are mostly automotive
examples from~\cite{HoxhaAF14arch1}---and compare performance between:
\begin{itemize}
\item our prototype, run for $\mathcal{M}$ and the original
  $\STL$ specification $\varphi$,\footnote{
    This is the control case of our experiments.  
    We do not use S-TaLiRo itself, because we would like to disregard
    the potential disadvantage caused by the communication between the
    $\AvSTL$ evaluator (the additional component) and S-TaLiRo. 
    We note that the $\AvSTL$ evaluator is capable of
    evaluating $\STL$ formulas, too.} and
\item our prototype, run for $\mathcal{M}$ and a
  \emph{refinement} of $\varphi$ given as an  $\AvSTL$ formula.
\end{itemize}
For benchmarks of a certain class we observe substantial performance
improvement:  sometimes the latter is several times faster; and in some
benchmarks we even see the latter succeed in falsification while the former 
fails to do so.

\noindent
\textit{Related Work}\;
Besides those which are discussed in the above
and the below, a closely related work
is~\cite{DBLP:journals/corr/AbbasHFDKU14} (its abstract appeared
in~\cite{DBLP:conf/iccps/AbbasHFDKU14}). There a notion of
\emph{conformance} between two models $\mathcal{M}_{1}$,
$\mathcal{M}_{2}$ is defined; and it is much like (an arity-2 variation
of) combination of  space and time
robustness. Its use in falsification and comparison with the current
approach is future work.

\noindent
\textit{Organization of the Paper}\;
In~\S{}\ref{sec:AvSTL} we introduce the logic $\AvSTL$: its syntax,
semantics, some basic properties and examples of temporal specifications
expressible in it. In~\S{}\ref{sec:algorithm}, building
on~\cite{DBLP:conf/cav/DonzeFM13}, an algorithm for computing
$\AvSTL$ truth values is introduced and its complexity is studied.
The algorithm is implemented and used
to enhance a falsification solver S-TaLiRo,
in~\S{}\ref{sec:experiments}, where experiment results are presented and
discussed.

We used colors in some figures for clarity. Consult the electronic
edition  in case the colors are unavailable.  Most of the proofs are
deferred to the appendix.

\textbf{Acknowledgments}
Thanks are due to
Georgios Fainekos, 
Tomoyuki Kaga,  Toshiki Kataoka, 
Hisashi Miyashita,
Kohei Suenaga and
Tomoya Yamaguchi
 for helpful
 discussions. The authors are supported by Grant-in-Aid for Young
 Scientists (A) No.\ 24680001, JSPS; and 
T.A.\ is supported by Grant-in-Aid for JSPS Fellows.

\section{Averaged Signal Temporal Logic $\AvSTL$}
\label{sec:AvSTL}
We introduce \emph{averaged STL} ($\AvSTL$). It is essentially an extension of 
$\textbf{MITL}$~\cite{DBLP:dblp_journals/jacm/AlurFH96} and
$\textbf{STL}$~\cite{DBLP:conf/formats/MalerN04} with so-called
\emph{averaged} temporal operators.  We describe its syntax and
its semantics (that is inspired by robust semantics
in~\cite{DBLP:journals/tcs/FainekosP09,DBLP:conf/formats/DonzeM10}).
We also exemplify the expressivity of the logic, by encoding 
common temporal specifications like expeditiousness,
persistence and  deadline.
Finally  we will
discuss the relationship to the previous robustness notions~\cite{DBLP:journals/tcs/FainekosP09,DBLP:conf/formats/DonzeM10} for $\STL$.

\subsection{Syntax}\label{subsec:syntax}
We let $\equiv$ stand for the syntactic equality.
We let  $\R$ denote the set of real numbers, with 
$\Rnn$ and $\Rnp$ denoting its obvious subsets.
We also fix the set $\Var$ of variables, each of which 
stands for a  physical quantity (velocity, temperature, etc.). 

\begin{mydefinition}[syntax]\label{def:syntax}
  In $\AvSTL$, the set $\AP$ of \emph{atomic propositions} and the set
  $\Fml$
  of \emph{formulas} are defined as follows.
    \[\small
    \begin{array}{rrl}
    \AP \ni 
      & l \,::=\,
      & x < r 
        \mid x \leq r 
        \mid x \geq r 
        \mid x > r 
        \quad \text{ where } x \in \Var, r \in \R\\
      \Fml \ni 
      &\varphi \,::=\,
      & \top
        \mid \bot
        \mid l 
        \mid \neg \varphi 
        \mid \varphi \vee \varphi 
        \mid \varphi \wedge \varphi 
        \mid \varphi \UntilOp{I} \varphi 
        \mid \varphi \TUntil{I} \varphi 
        \mid \varphi \Release{I} \varphi 
        \mid \varphi \TRelease{I} \varphi
    \end{array}
  \]
  Here $I$ is a closed non-singular interval in $\Rnn$,
  i.e. $I=[a,b]$ or $[a, \infty)$ where $a<b$.
  The overlined operator
  $\TUntil{I}$ is called the \emph{averaged-until} operator.
  
  We introduce the following connectives as abbreviations, as usual:
  \begin{math}
    \varphi_1 \to \varphi_2  \equiv  (\neg \varphi_1) \vee \varphi_2
  \end{math},
  \begin{math}
    \DiaOp{I} \varphi  \equiv  \top \UntilOp{I} \varphi
  \end{math},
  \begin{math}
    \BoxOp{I} \varphi  \equiv  \bot \Release{I} \varphi 
  \end{math},
  \begin{math}
    \TDiaOp{I} \varphi  \equiv  \top \TUntil{I} \varphi
  \end{math} 
  and
  \begin{math}
    \TBoxOp{I} \varphi  \equiv  \bot \TRelease{I} \varphi 
  \end{math}.
  %
  We omit subscripts $I$ for temporal operators if $I = [0, \infty)$.
  The operators  $\TRelease{I}$, $\TDiaOp{I}$ and $\TBoxOp{I}$ are called the
  \emph{averaged-release}, \emph{averaged-eventually} and \emph{averaged-henceforth} operators, respectively.
  We say a formula $\varphi$ is \emph{averaging-free}
  if it does not contain any averaged temporal operator.

  \auxproof{
    As usual, we can exploit de Morgan-like dualities and push negations
    inwards towards atomic propositions, leading to 
    \emph{negation normal forms (NNF)}. 
    Explicitly, the set $\FmlNNF$ of NNF formulas is defined as
    follows, where $l \in \AP$.
    \[
      \begin{array}{rll}
        \FmlNNF \ni \varphi &\;::=\;&
        \infty \mid \neg \infty 
        \mid l \mid \neg l \mid \varphi_1 \vee \varphi_2 
        \mid \varphi_1 \wedge \varphi_2 \mid\\
        &&\varphi_1 \UntilOp{I} \varphi_2 \mid \varphi_1 \TUntil{I} \varphi_2
        \mid \varphi_1 \Release{I} \varphi_2 \mid \varphi_1 \TRelease{I} \varphi_2
      \end{array}
    \]
  }
\end{mydefinition}

\subsection{Robust Semantics}\label{subsec:semantics}
 $\AvSTL$ formulas, much like  $\STL$ formulas in~\cite{DBLP:journals/tcs/FainekosP09,DBLP:conf/formats/DonzeM10}, are interpreted over 
(real-valued, continuous-time) \emph{signals}. The latter stand for
trajectories of hybrid systems.
\begin{mydefinition}[signal]\label{def:signal}
  A \emph{signal} over $\Var$ is a function $\sigma\colon
  \Rnn\to (\R^{\Var})$; it is therefore a bunch of physical quantities indexed
  by a continuous notion of time.

For a signal $\sigma$  and $t\in \Rnn$,
 $\sigma^t$ denotes the \emph{$t$-shift} of $\sigma$, that is,
$\sigma^t(t') \Defeq \sigma(t+t')$.

\end{mydefinition}

\begin{wrapfigure}[5]{r}{0pt}
\centering
\includegraphics[trim=0cm 0cm 0cm 3cm, width=.33\textwidth]{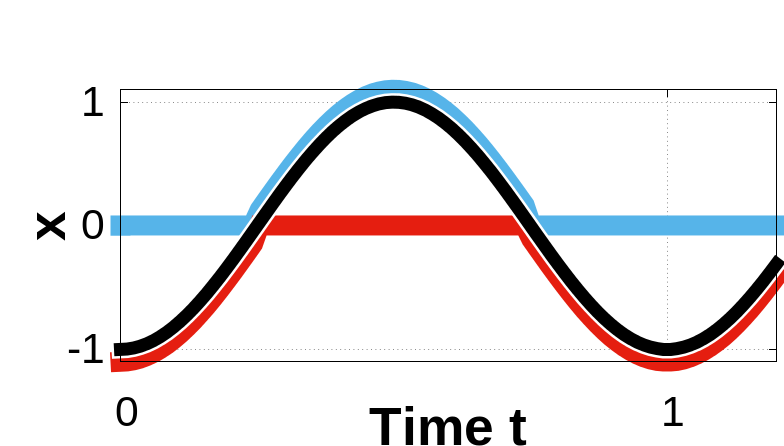}
\end{wrapfigure}
The interpretation of a formula $\varphi$ over a signal $\sigma$ is given by two different 
``truth values,'' namely \emph{positive} and \emph{negative robustness}.
They are denoted by $      \Robust{\sigma}{\varphi}^{+}$ and 
 $      \Robust{\sigma}{\varphi}^{-}$, respectively. 

We will always have
\begin{math}
 \Robust{\sigma}{\varphi}^{+} \ge 0
\end{math} and
\begin{math}
 \Robust{\sigma}{\varphi}^{-} \le 0
\end{math}.
We will also see that, for averaging-free $\varphi$,
it is never the case that
\begin{math}
  \Robust{\sigma}{\varphi}^{+} > 0
\end{math}
and
\begin{math}
 \Robust{\sigma}{\varphi}^{-} <0
\end{math}
hold at the same time. 
See the figure on the right for an example, where a sine-like (black) curve is
a signal $\sigma$. The blue and red curves stand for the positive and
negative robustness, of the formula $x\ge 0$ over the ($t$-shifted) signal
$\sigma^{t}$, respectively.

\begin{mydefinition}[positive/negative robustness]\label{def:semantics}
  Let $\sigma \colon \Rnn \to \R^\Var$ be a signal 
  and $\varphi$ be an $\AvSTL$ formula.
  We define the \emph{positive robustness} 
  $\Robust{\sigma}{\varphi}^{+} \in \Rnn \cup \{\infty\}$ 
  and the \emph{negative  robustness} 
  $\Robust{\sigma}{\varphi}^{-} \in \Rnp \cup \{ - \infty\}$ 
  by mutual induction, as shown in Table~\ref{table:robustness}. 
  Here $\sqcap$ and $\sqcup$ denote infimums and supremums of real numbers, respectively.
  \begin{table}[t]
 \begin{tabular}{l}
     \scalebox{0.86}{
      \begin{math}
      \begin{array}{rll}
        \Robust{\sigma}{\top}^{+} & \Defeq & \infty\\
        \Robust{\sigma}{\bot}^{+} & \Defeq & 0\\
        \Robust{\sigma}{x < r}^{+} & \Defeq & 0\sqcup (r - \sigma(0)(x))\\
        \Robust{\sigma}{x \leq r}^{+} & \Defeq & 0\sqcup(r - \sigma(0)(x))\\
        \Robust{\sigma}{x \geq r}^{+} & \Defeq & 0\sqcup(\sigma(0)(x) - r)\\
        \Robust{\sigma}{x > r}^{+} & \Defeq & 0\sqcup( \sigma(0)(x) - r)\\
        \Robust{\sigma}{\neg \varphi}^{+} & \Defeq & - \Robust{\sigma}{\varphi}^{-}\\
        \Robust{\sigma}{\varphi_1 \vee \varphi_2}^{+} & \Defeq & 
        \Robust{\sigma}{\varphi_1}^{+} \sqcup \Robust{\sigma}{\varphi_2}^{+}\\
        \Robust{\sigma}{\varphi_1 \wedge \varphi_2}^{+} & \Defeq & 
        \Robust{\sigma}{\varphi_1}^{+} \sqcap \Robust{\sigma}{\varphi_2}^{+}\\
        \\
      \end{array}
      \quad
      \begin{array}{l}
        \begin{array}{rll}
          \Robust{\sigma}{\varphi_1 \UntilOp{I} \varphi_2}^{+} 
          & \Defeq 
          & \Vee{t \in I} 
            (\Robust{\sigma^t}{\varphi_2}^{+} \sqcap 
            \Wedge{t' \in [0, t)} \Robust{\sigma^{t'}}{\varphi_1}^{+})\\
          \Robust{\sigma}{\varphi_1 \Release{I} \varphi_2}^{+} 
          & \Defeq 
          & \Wedge{t \in I} 
            (\Robust{\sigma^t}{\varphi_2}^{+} \sqcup 
            \Vee{t' \in [0, t)} \Robust{\sigma^{t'}}{\varphi_1}^{+})\\
        \end{array}
        \\
        \begin{array}{l}
          \Robust{\sigma}{\varphi_1 \TUntil{I} \varphi_2}^{+} \Defeq\\
          \quad
          \begin{cases}
            \Frac{1}{b - a} \Int_{a}^{b} 
            \Robust{\sigma}{\varphi_1 \UntilOp{I \cap [0, \tau]} \varphi_2}^{+} d\tau 
            & \text{($I$ is bounded)}\\
            \Robust{\sigma}{\varphi_1 \UntilOp{I} \varphi_2}^{+} 
            & \text{($I$ is unbounded)}\\
          \end{cases}
	  \\
          \Robust{\sigma}{\varphi_1 \TRelease{I} \varphi_2}^{+} \Defeq\\
          \quad
          \begin{cases}
            \Frac{1}{b - a} \Int_{a}^{b} 
            \Robust{\sigma}{\varphi_1 \Release{I \cap [0, \tau]} \varphi_2}^{+} d\tau 
            & \text{($I$ is bounded)}\\
            \Robust{\sigma}{\varphi_1 \Release{I} \varphi_2}^{+} 
            & \text{($I$ is unbounded)}\\
          \end{cases}\\
        \end{array}
      \end{array}
    \end{math}
   }
   \\
   \hline

    \scalebox{0.86}{
      \begin{math}
        \begin{array}{rll}
        \Robust{\sigma}{\top}^{-} & \Defeq & 0\\
        \Robust{\sigma}{\bot}^{-} & \Defeq & - \infty\\
        \Robust{\sigma}{x < r}^{-} & \Defeq & 0\sqcap( r - \sigma(0)(x))\\
        \Robust{\sigma}{x \leq r}^{-} & \Defeq & 0\sqcap( r - \sigma(0)(x))\\
        \Robust{\sigma}{x \geq r}^{-} & \Defeq & 0\sqcap( \sigma(0)(x) - r)\\
        \Robust{\sigma}{x > r}^{-} & \Defeq & 0\sqcap( \sigma(0)(x) - r)\\
        \Robust{\sigma}{\neg \varphi}^{-} & \Defeq & - \Robust{\sigma}{\varphi}^{+}\\
        \Robust{\sigma}{\varphi_1 \vee \varphi_2}^{-} & \Defeq & 
        \Robust{\sigma}{\varphi_1}^{-} \sqcup \Robust{\sigma}{\varphi_2}^{-}\\
        \Robust{\sigma}{\varphi_1 \wedge \varphi_2}^{-} & \Defeq & 
        \Robust{\sigma}{\varphi_1}^{-} \sqcap \Robust{\sigma}{\varphi_2}^{-}\\
          \\
      \end{array}
      \quad
      \begin{array}{l}
        \begin{array}{rl}
          \Robust{\sigma}{\varphi_1 \UntilOp{I} \varphi_2}^{-} 
          & \Defeq 
            \Vee{t \in I} 
            (\Robust{\sigma^t}{\varphi_2}^{-} \sqcap \Wedge{t' \in [0, t)} \Robust{\sigma^{t'}}{\varphi_1}^{-})\\
          \Robust{\sigma}{\varphi_1 \Release{I} \varphi_2}^{-} 
          & \Defeq  
            \Wedge{t \in I} 
            (\Robust{\sigma^t}{\varphi_2}^{-} \sqcup \Vee{t' \in [0, t)} \Robust{\sigma^{t'}}{\varphi_1}^{-})\\
        \end{array}\\
        \begin{array}{l}
          \Robust{\sigma}{\varphi_1 \TUntil{I} \varphi_2}^{-} \Defeq\\
          \quad
          \begin{cases}
            \Frac{1}{b - a} \Int_{a}^{b} \Robust{\sigma}{\varphi_1 \UntilOp{I \cap [0, \tau]} \varphi_2}^{-} d\tau 
            & \text{($I$ is bounded)}\\
            \Robust{\sigma}{\varphi_1 \UntilOp{I} \varphi_2}^{-} &
            \text{($I$ is unbounded)}\\
          \end{cases}\\
          \Robust{\sigma}{\varphi_1 \TRelease{I} \varphi_2}^{-}  \Defeq\\
          \quad
          \begin{cases}
            \Frac{1}{b - a} \Int_{a}^{b} \Robust{\sigma}{\varphi_1 \Release{I \cap [0, \tau]} \varphi_2}^{-} d\tau 
            & \text{($I$ is bounded)}\\
            \Robust{\sigma}{\varphi_1 \Release{I} \varphi_2}^{-} &
            \text{($I$ is unbounded)}\\
          \end{cases}\\
        \end{array}
      \end{array}
    \end{math}
    }
 \end{tabular}
 \caption{Definition of positive and negative robustness}
 \label{table:robustness}
\end{table}
\end{mydefinition}
The definition in Table~\ref{table:robustness} is much like the one for
$\STL$~\cite{DBLP:conf/formats/DonzeM10,DBLP:conf/cav/DonzeFM13},\footnote{
  There is no distinction between strict inequalities ($<$) and
  non-strict ones ($\le$). This is inevitable in the current
  robustness framework. This is also the case with $\STL$
  in~\cite{DBLP:conf/formats/DonzeM10,DBLP:conf/cav/DonzeFM13}. }  
except for the averaged modalities on which a detailed account follows shortly.
Conjunctions and
disjunctions are interpreted by infimums and supremums, in a
straightforward manner.

\begin{wrapfigure}[8]{r}{0pt}
  \begin{minipage}[r]{.4\textwidth}
  \centering
  \includegraphics[trim=0cm 2cm 0cm 2cm, width=0.9\textwidth,natwidth=640,natheight=384]{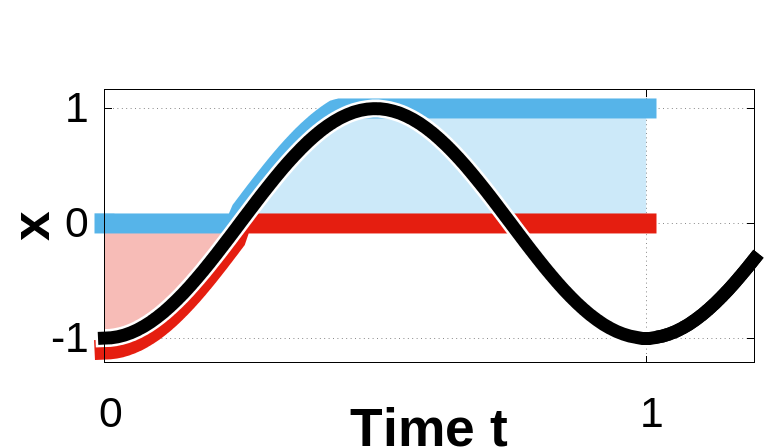}
  \caption{
    The positive and negative robustness of 
    $\TDiaOp{[0, 1]} (x \geq 0)$ at $t=0$.}
  \label{fig:averagedDiamond}
    
  \end{minipage}
\end{wrapfigure}

Fig.~\ref{fig:averagedDiamond} illustrates the semantics of 
averaged-temporal operators---the novelty of our logic $\AvSTL$. Specifically, 
the black line designates a
signal $\sigma$ whose only variable is $x$; and we consider the ``averaged-eventually''
formula $\TDiaOp{[0, 1]} (x \geq 0)$. For this formula, the definition in
Table~\ref{table:robustness} specializes to:
\[
  \begin{array}{l}
    \Robust{\sigma}{\TDiaOp{[0, 1]} (x \geq 0)}^{+} \\
    =
    \Int_{0}^{1} \Bigl(\,\Vee{\tau' \in [0 , \tau]} 0\sqcup\,
    \sigma(\tau')(x)\,\Bigr) \,d\tau\enspace, 
  \end{array}
  \qquad\text{and}\qquad
  \begin{array}{l}
    \Robust{\sigma}{\TDiaOp{[0, 1]} (x \geq 0)}^{-}
    \\
    =
    \Int_{0}^{1}\Bigl(\, \Vee{\tau' \in [0 , \tau]} 0
    \sqcap\,\sigma(\tau')(x) 
    \,\Bigr)\,d\tau\enspace .
  \end{array}
\]
These values obviously coincide with the sizes of the blue and red
areas in Fig.~\ref{fig:averagedDiamond}, respectively. Through this ``area''
illustration 
of the averaged-eventually operator 
we see that: the sooner
$\varphi$ is true, the more (positively) robust $\TDiaOp{I}\varphi$ is.
It is also clear from Fig.~\ref{fig:averagedDiamond} that our semantics captures space robustness
too: the bigger a vertical margin is, the bigger an area is.

\begin{myremark}\label{rem:separationPosNegRobustness}
Presence of averaged temporal operators forces separation of two
 robustness measures (positive and negative). Assume otherwise, i.e.
 that we have one robustness measure
that can take both positive and negative values; then robustness that floats  between
positive and negative values over time
can ``cancel out'' after an average is taken. This leads to the failure
of \emph{soundness} (see Prop.~\ref{prop:diaRefinement} and \ref{prop:BoxRefinement};
also~\cite{DBLP:journals/tcs/FainekosP09,DBLP:conf/formats/DonzeM10}),
and then a positive robustness value no longer witnesses the Boolean truth
of  (the qualitative variant of) the formula. This is not convenient in
 the application to falsification.
\end{myremark}


\auxproof{
 \begin{myremark}
 In Def.~\ref{def:semantics},
 we define the value of positive/negative temporal robustness
 of averaged- modalities
 as a definite integral. 
 From the following Lem.~\ref{lemma:untilIsMonotone},
 we can check the integrability
 of them;
 all 
 $\Robust{\sigma}{\varphi_1 \UntilOp{[t,\tau]} \varphi_2}^{+}$,
 $\Robust{\sigma}{\varphi_1 \UntilOp{[t,\tau]} \varphi_2}^{-}$,
 $\Robust{\sigma}{\varphi_1 \Release{[t,\tau]} \varphi_2}^{+}$, and
 $\Robust{\sigma}{\varphi_1 \Release{[t,\tau]} \varphi_2}^{-}$ 
 are monotonically increasing or decreasing
 over $\tau \in [t, t']$,
 hence these functions are integrable.
 \end{myremark}
}

\subsection{Basic Properties of $\AvSTL$}
\begin{mylemma}[temporal monotonicity]\label{lemma:untilIsMonotone}
  Let 
  $0 \leq t_0 < t \leq t'$. 
  The following hold.
   \[\small
    \begin{array}{rclrcl}
      \Robust{\sigma}{\varphi_1 \UntilOp{[t_0,t]} \varphi_2}^{+} 
      &\leq &
      \Robust{\sigma}{\varphi_1 \UntilOp{[t_0,t']} \varphi_2}^{+}
      \quad
      &
      \Robust{\sigma}{\varphi_1 \UntilOp{[t_0,t]} \varphi_2}^{-} 
      &\leq &
      \Robust{\sigma}{\varphi_1 \UntilOp{[t_0,t']} \varphi_2}^{-}
      \\
      \Robust{\sigma}{\varphi_1 \Release{[t_0,t]} \varphi_2}^{+} 
      &\geq&
      \Robust{\sigma}{\varphi_1 \Release{[t_0,t']} \varphi_2}^{+}
      &
      \Robust{\sigma}{\varphi_1 \Release{[t_0,t]} \varphi_2}^{-} 
      &\geq&
      \Robust{\sigma}{\varphi_1 \Release{[t_0,t']} \varphi_2}^{-}
    \end{array}
  \] 
  The inequalities hold also for the averaged temporal operators. \myqed                        
\end{mylemma}
We can now see well-definedness of Def.~\ref{def:semantics}:
we need that the integrals are defined; and the lemma shows that
the integrated functions are monotone, hence Riemann integrable.

  In Def.~\ref{def:semantics},
 the definitions for averaged operators with an infinite endpoint
 (like $\TUntil{[0, \infty)}{\varphi}$) are given in terms of non-averaged
 operators. This is so that their well-definedness is immediate; the
 following lemma justifies those definitions.
%
\begin{mylemma}\label{lem:CorrespondenceBetweenUntilAndTUntil}
  For any $t_0 \in \Rnn$, 
  \begin{math}
    \Robust{\sigma}{\varphi_1 \UntilOp{[t_0, \infty)} \varphi_2}^{+} 
      =
      \Lim{t \to \infty}
      \Robust{\sigma}{\varphi_1 \TUntil{[t_0, t]} \varphi_2}^{+}
  \end{math}. 
  The same is true if we replace $\sem{\place}^{+}$ with 
$\sem{\place}^{-}$, and if we replace $\UntilOp{}$ with $\Release{}$.
\myqed
\end{mylemma}

\auxproof{

\begin{mylemma}[logical monotonicity]
  \label{lem:monotonicity}
  Let $\C$ be a \emph{positive context}, that is, an $\AvSTL$ formula 
  with a hole $[\,]$ at a positive position. We have
  \begin{align*}
    \begin{array}{rcl}
      \forall \sigma.\; \Robust{\sigma}{\varphi}^{+} 
      \leq \Robust{\sigma}{\varphi'}^{+}
      \quad&\text{implies}
      &\quad 
        \forall \sigma.\; \Robust{\sigma}{\C[\varphi]}^{+} 
        \leq
        \Robust{\sigma}{\C[\varphi']}^{+}\enspace;\;\text{and}\\
      \forall \sigma.\; \Robust{\sigma}{\varphi}^{-} 
      \leq \Robust{\sigma}{\varphi'}^{-}
      \quad&\text{implies}
      &\quad 
        \forall \sigma.\; \Robust{\sigma}{\C[\varphi]}^{-} 
        \leq 
        \Robust{\sigma}{\C[\varphi']}^{-} \enspace.
    \end{array}  
  \end{align*}
\end{mylemma}
\begin{myproof}
  By induction on the construction of the positive context $\C$; the
  latter is thought of as a formula of the negative normal form with a
  hole.  \myqed
\end{myproof}

}



\subsection{Common Temporal Specifications Expressed in $\AvSTL$}\label{subsec:examplesExpressivity}
Here we shall exemplify the expressivity of $\AvSTL$, by encoding 
typical temporal specifications  encountered in the model-based
development of cyber-physical systems.

\begin{myremark}
\label{rem:propVar}
 In what follows we sometimes use \emph{propositional variables} such as
 $\mathtt{airbag}$
 and $\mathtt{gear}_{i}$. For example,  $\mathtt{gear}_{2}$
 is a shorthand for the atomic formula $x_{\mathtt{gear}_{2}} \ge 0$ in $\AvSTL$, where
 the variable $x_{\mathtt{gear}_{2}}$ is assumed to take a discrete value
 ($1$ or $-1$).
\end{myremark}


\subsubsection{Expeditiousness ($\TDiaOp{I}\varphi$)}
Consider the following informal specification: 
\emph{after heavy braking,
  the airbag must operate
  within 10 ms.} 
Its formalization in $\STL$ is straightforward by the formula 
\begin{math}
  \BoxOp{} (
  \mathtt{heavyBraking} \to 
  \DiaOp{[0,10]}\mathtt{airbag} )
\end{math}. 
However, an airbag that operates
after 1 ms.\ is naturally more desirable than one that operates
after 9.99 ms. The $\STL$ formula fails to discriminate between these
two airbags.

\begin{figure}[tbp]
   \begin{minipage}{.33\textwidth}
        \includegraphics[width=\textwidth,natwidth=640,
    natheight=384]{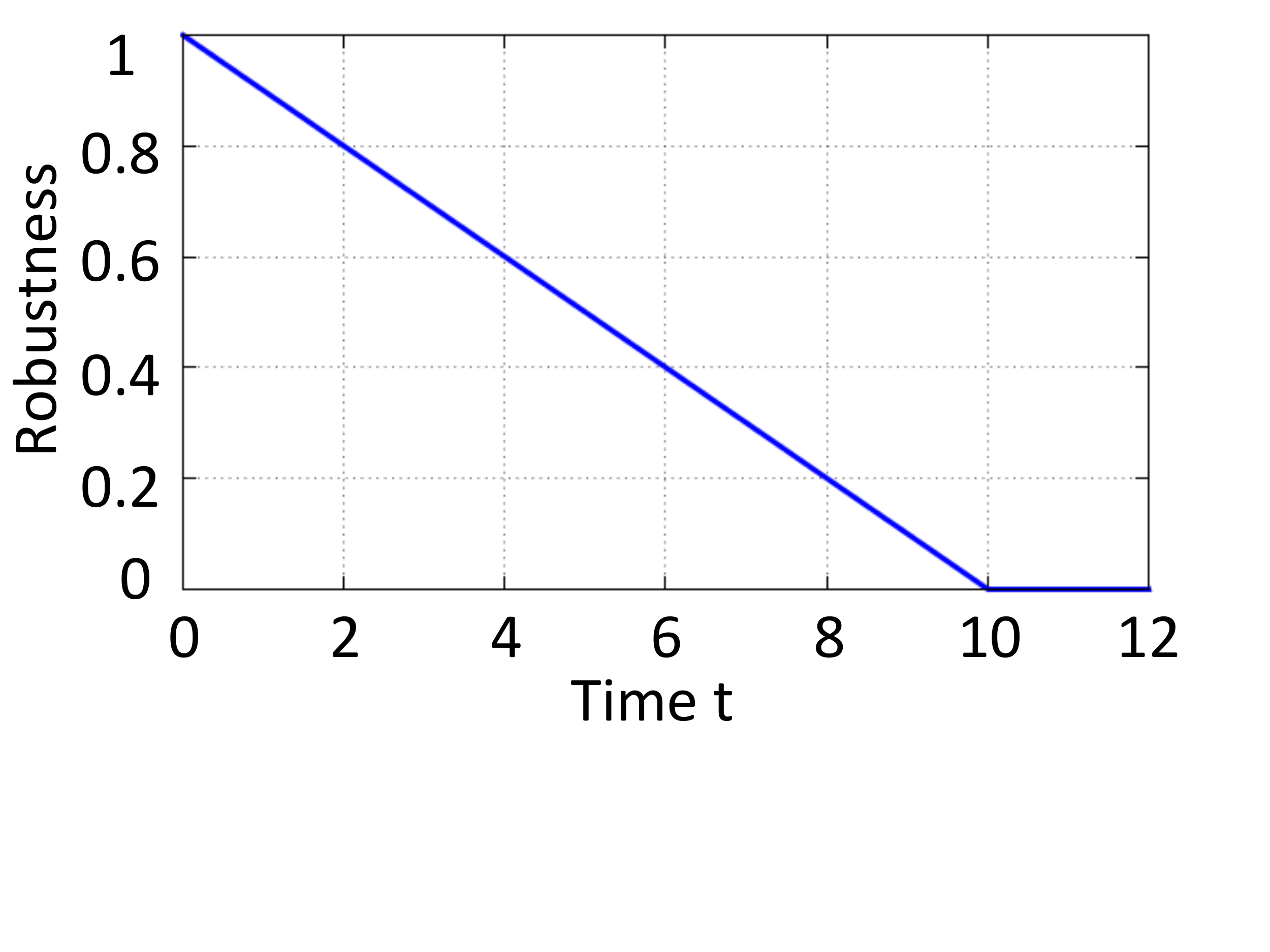}

    \vspace{-3.5em}
      \caption{Expeditiousness}
  \label{fig:expeditiousness}
   \end{minipage}
   \begin{minipage}{.33\textwidth}
        \includegraphics[width=\textwidth,natwidth=640,
    natheight=384]{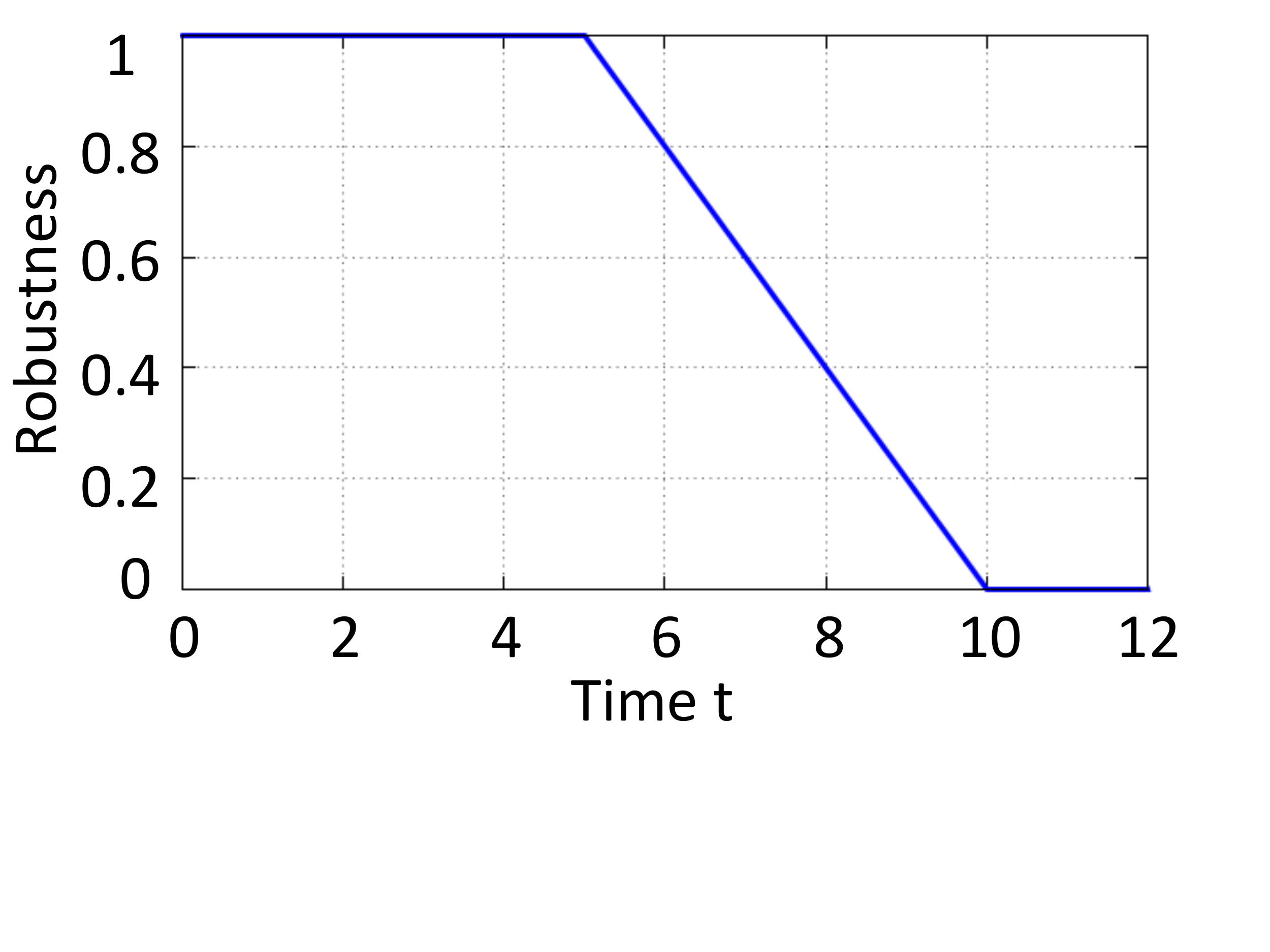}

        \vspace{-3.5em}
      \caption{Deadline}
  \label{fig:deadline}
   \end{minipage}
    \begin{minipage}{.33\textwidth}
        \includegraphics[width=\textwidth,natwidth=640,
     natheight=384]{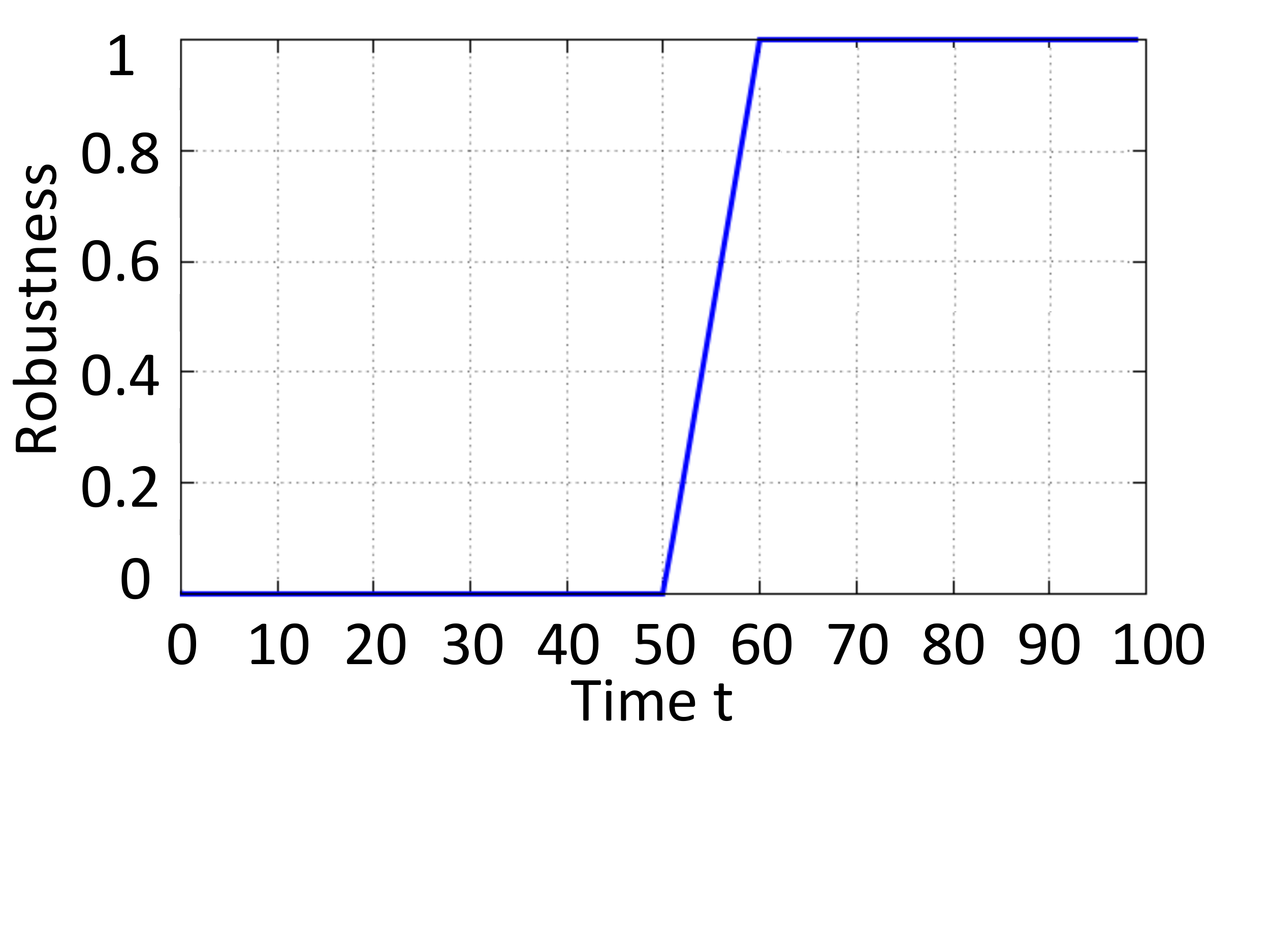}

         \vspace{-3.5em}
      \caption{Persistence}
  \label{fig:persistence}
   \end{minipage}
\end{figure}

Such  \emph{expeditiousness} (``as soon as possible'') requirements
are more adequately  modeled in $\AvSTL$,  using the averaged-eventually modality
 $\TDiaOp{I}$. See Fig.~\ref{fig:expeditiousness}, where
the horizontal axis is
for time $t$. The vertical axis in the figure stands for the positive
robustness  value
$\Robust{\sigma_{t}}{\TDiaOp{[0,10]}\mathtt{airbag}}^{+}$ of the formula
$\TDiaOp{[0,10]}\mathtt{airbag}$, where $\sigma_{t}$ is a signal in
which $\mathtt{airbag}$ operates 
(i.e. $x_{\mathtt{airbag}}$ becomes from $-1$ to $1$) at time $t$. 
We see that the formula successfully distinguishes an early-bird
airbag from a lazy one.

Therefore the $\AvSTL$ formula
\begin{math}
     \BoxOp{} (
    \mathtt{heavyBraking} \to 
\TDiaOp{[0,10]}\mathtt{airbag} )
\end{math} formalizes
 a (refined) informal specification that: after heavy braking, the airbag must
operate within 10 ms; \emph{but the sooner the better}. It is not hard
to expect that the latter is
 more faithful to the designer's intention than the
original informal specification.

\subsubsection{Deadline ($\DiaOp{[0,T]}\varphi\lor
   \TDiaOp{[T,T+\delta]}\varphi$)}
 The expeditiousness-type requirement that we have discussed is sometimes too
 strict. Let us consider the following scenario: there is a deadline set
 at time $T$ and arrival by then is rewarded no matter how late; and then there is a
 deadline extension by time $\delta$ and arrival between the deadline
 and the extended one is rewarded too, but with certain deduction. 

Such
a \emph{deadline} specification is expressed in $\AvSTL$ by the formula 
$\DiaOp{[0,T]}\varphi \lor \TDiaOp{[T,T+\delta]}\varphi$, 
combining non-averaged and averaged
eventually modalities. See Fig.~\ref{fig:deadline}, where the
positive robustness of the formula
\begin{math}
 (\DiaOp{[0,5]}\mathtt{airbag})\lor
   (\TDiaOp{[5,5+5]}\mathtt{airbag})
\end{math}
is plotted, for the same signals $\sigma_{t}$ as before 
(i.e. in $\sigma_{t}$ the airbag operates at time $t$).

\subsubsection{Persistence
   ($\BoxOp{[0,T]}\varphi\land\TBoxOp{[T,T+\delta]}\varphi$)}
\emph{Persistence} (``for as long as possible'') specifications are dual
to deadline ones and expressed by a formula
$\BoxOp{[0,T]}\varphi\land\TBoxOp{[T,T+\delta]}\varphi$. An example is
the following informal specification on automatic transmission:
\emph{when a gear shifts into first,
it never shifts into any other gear for the coming 50 ms.} A likely
intention behind it is to prevent mechanical wear of gears that is caused by
frequent gear shifts. In this case the following specification 
would be more faithful to the intention: when a gear shifts into first,
it never shifts into any other gear for the coming 50 ms., \emph{and
preferably for longer}. This is formalized by the formula
\begin{math}
    \BoxOp{} 
    (
    \mathtt{shiftIntoGear_1} \to
\BoxOp{[0,50]}\mathtt{gear_1}\land\TBoxOp{[50,50+\delta]}\mathtt{gear_1}
    )
\end{math}.

For illustration, Fig.~\ref{fig:persistence} plots
the positive robustness of
$\BoxOp{[0,50]}\mathtt{gear_1}\land\TBoxOp{[50,60]}\mathtt{gear_1}$
for signals $\sigma'_{t}$, where $\mathtt{gear_1}$ is true  in
$\sigma'_{t}$ from time $0$ to $t$, and is false afterwards.

\subsubsection{Other Temporal Specifications}
Expressivity of $\AvSTL$ goes beyond the three examples that we have
seen---especially after the extension of the language with
\emph{time-reversed} averaged temporal operators. The reversal of time
here corresponds to the symmetry between \emph{left} and \emph{right}
time robustness in~\cite{DBLP:conf/formats/DonzeM10}.  Such an
extension of $\AvSTL$ enables us to express specifications like
\emph{punctuality}
 (``no sooner, no later'') and \emph{periodicity}. The details will be
 reported in another venue.

\subsection{Soundness of  Refinements from $\STL$ to $\AvSTL$}\label{subsec:soundnessOfEnrichments}
In~\S\ref{subsec:examplesExpressivity} we have seen some scenarios where
an $\STL$ specification is \emph{refined} into an $\AvSTL$ one so that
it more faithfully reflects the designer's intention. The following two
are prototypical:
\begin{itemize}
 \item{}
  \textbf{($\Diamond$-refinement)}
  the refinement
 of $\DiaOp{I}\varphi$ (``eventually $\varphi$, within $I$'') into
 $\TDiaOp{I}\varphi$
 (``eventually $\varphi$ within $I$, but as soon as possible''); and
 \item{}
  \textbf{($\Box$-refinement)}
 the refinement of
 $\BoxOp{[a,b]}\varphi$ 
 (``always $\varphi$ throughout $[a,b]$'') into
 $\BoxOp{[a,b]}\varphi\land\TBoxOp{[b,b+\delta]}\varphi$ (``always
 $\varphi$ throughout $[a,b]$, and desirably also in $[b,b+\delta]$'').
\end{itemize} 
The following \emph{soundness} 
results
guarantee validity of the use of these refinements in falsification
problems. \emph{Completeness}, in a suitable sense, holds too.

\begin{mydefinition}
  \label{def:context}
  A \emph{positive context} is
  an $\AvSTL$ formula with a hole $[\,]$ at a positive position.
  Formally, 
  the set of positive contexts is defined as follows:
  \[
  \begin{array}{rl}
    \C \,::=\,& [\,] \mid \C \vee \varphi \mid \varphi \vee \C 
                \mid \C \wedge \varphi \mid \varphi \wedge \C
                \mid \C \UntilOp{I} \varphi 
                \mid \varphi \UntilOp{I} \C 
                \mid \C \TUntil{I} \varphi 
                \mid \varphi \TUntil{I} \C\\
              & \mid \C \Release{I} \varphi 
                \mid \varphi \Release{I} \C 
                \mid \C \TRelease{I} \varphi
                \mid \varphi \TRelease{I} \C \quad
                \text{ where $\varphi$ is an $\AvSTL$ formula. }
  \end{array}
  \]
  For a positive context $\C$ and an $\AvSTL$ formula $\psi$,
  $\C[\psi]$ denotes the formula 
  obtained by substitution of $\psi$
  for the hole $[\,]$ in $\C$.
\end{mydefinition}

\noindent
\begin{minipage}{\textwidth}
 \begin{myproposition}[soundness and completeness of $\DiaOp{}$-refinement]
 \label{prop:diaBoxReplacement}
 \label{prop:diaRefinement}
  Let $\C$ be a positive context.
  Then
  \begin{math}
    \Robust{\sigma}{\C[\TDiaOp{[a,b]}\varphi]}^{+} > 0
  \end{math}
  implies
  \begin{math}
    \Robust{\sigma}{\C[\DiaOp{[a,b]}\varphi]}^{+} > 0.
  \end{math}
  Moreover, for any $b'$ such that
  \begin{math}
    b' < b
  \end{math},
  \begin{math}
    \Robust{\sigma}{\C[\DiaOp{[a,b']}\varphi]}^{+} > 0
  \end{math}
  implies
  \begin{math}
    \Robust{\sigma}{\C[\TDiaOp{[a,b]}\varphi]}^{+} > 0
  \end{math}
  \myqed
 \end{myproposition}
\end{minipage}

\noindent
\begin{minipage}{\textwidth}
\begin{myproposition}[soundness and completeness of $\BoxOp{}$-refinement]
\label{prop:BoxRefinement}
  Let $\C$ be a positive context.
  Then
  \begin{math} 
    \Robust{\sigma}{\C[\BoxOp{[a,b]}\varphi \wedge \TBoxOp{[b,b+\delta]}\varphi]}^{+} > 0
  \end{math}
  implies
\begin{math}
    \Robust{\sigma}{\C[\BoxOp{[a,b]}\varphi]}^{+} > 0 
  \end{math}. 
 Moreover, for any $b'>b$, 
  \begin{math}
    \Robust{\sigma}{\C[\BoxOp{[a,b']}\varphi]}^{+} > 0 
  \end{math} implies
\begin{math}
    \Robust{\sigma}{\C[\BoxOp{[a,b]}\varphi \wedge \TBoxOp{[b,b+\delta]}\varphi]}^{+} > 0
  \end{math}.
  \myqed
\end{myproposition}
\end{minipage}

　

\subsection{Relationship to Previous Robustness Notions}
Our logic $\AvSTL$ captures 
\emph{space robustness}~\cite{DBLP:journals/tcs/FainekosP09}---the
first robustness notion  proposed for $\textbf{MITL}$/$\STL$,
see~\S\ref{sec:introduction}---because the averaging-free fragment of $\AvSTL$ coincides with
$\STL$ and its space robust semantics, modulo the separation of positive
and negative robustness (Rem.~\ref{rem:separationPosNegRobustness}).
\auxproof{
More precisely,
the positive robustness of an averaged-free $\AvSTL$ formula $\varphi$
corresponds to 
the space robustness in $\STL$
if $\varphi$ is (qualitatively) true,
and so does the negative one if $\varphi$ is false.
}

\begin{wrapfigure}[7]{r}{0pt}
\centering
        \includegraphics[clip,trim=0cm 4cm 0cm
 0cm,width=.35\textwidth]{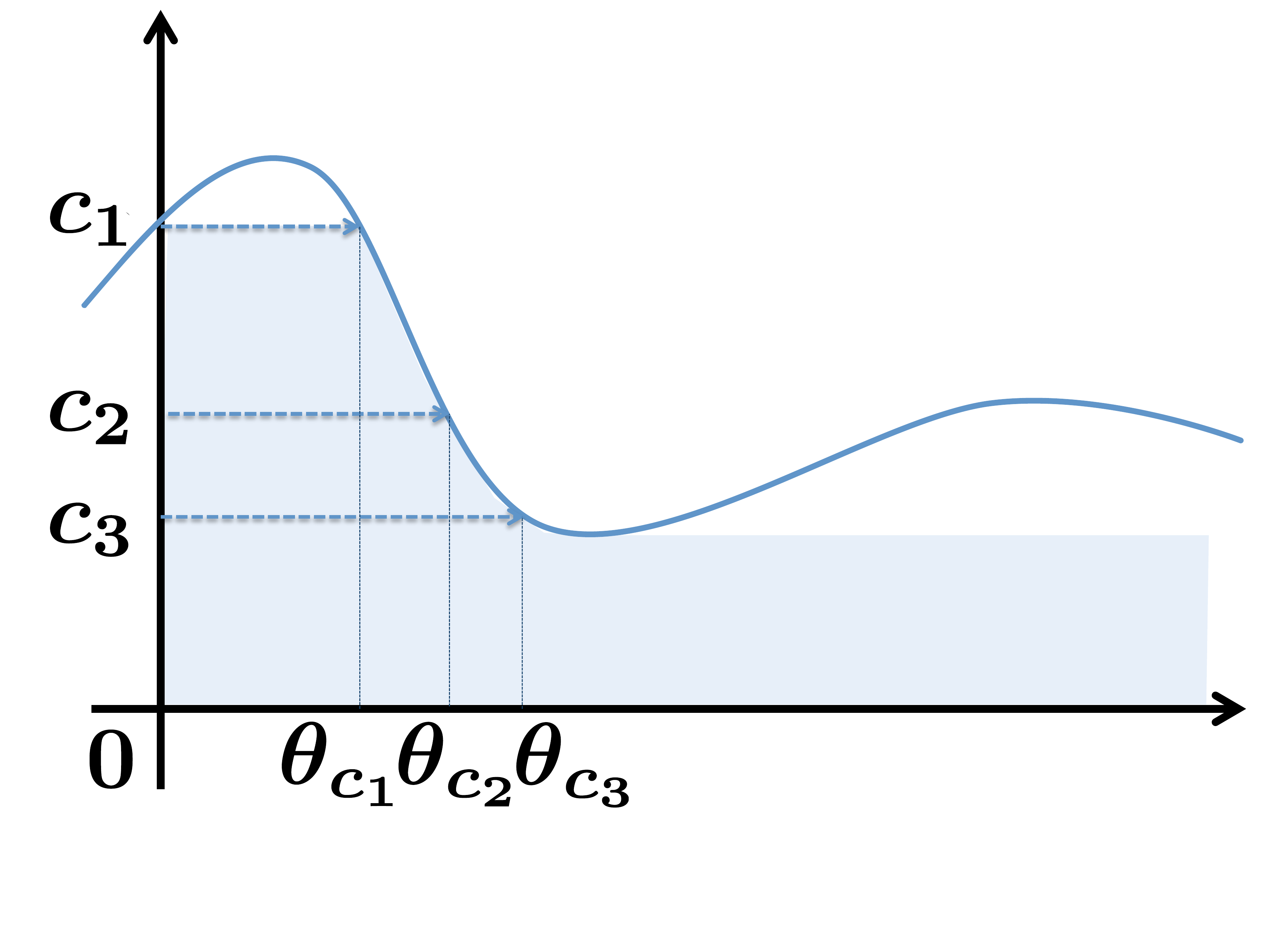}
\end{wrapfigure}
The relationship to \emph{space-time robustness} proposed
in~\cite{DBLP:conf/formats/DonzeM10} is
interesting. In~\cite{DBLP:conf/formats/DonzeM10} they combine time and
space robustness in the following way: for each time $t$ and each space robustness value $c>0$,
\emph{(right) time robustness relative to $c$}, denoted by
$\theta^{+}_{c}(\varphi,\sigma,t)$, is defined by ``how long
after time $t$ the formula $\varphi$
maintains space robustness $c$.'' See the figure
on the right, where the space-time robustness
$\theta^{+}_{c}(x\ge 0,\sigma,0)$ is depicted.

After all, space-time robustness in~\cite{DBLP:conf/formats/DonzeM10} is a function from $c$ to 
$\theta^{+}_{c}(\varphi,\sigma,t)$; and one would like some real number as
its characteristic. A natural choice of such is the \emph{area}
surrounded 
by
the graph of the function (the shaded area in the figure), and it is
 computed in the same way as
\emph{Lebesgue integration}, as the figure suggests. 

What corresponds in our $\AvSTL$
framework to this ``area'' characteristic value  is the robustness of the formula
$\TBoxOp{[0,\infty)}(x\ge 0)$ computed by Riemann
integration (here we have to
ignore the normalizing factor $\frac{1}{b-a}$ in
Table~\ref{table:robustness}). Therefore,  very roughly speaking:
our ``averaged'' robustness is a real-number characteristic value of  the
space-time robustness
in~\cite{DBLP:conf/formats/DonzeM10}; and the
correspondence
is via the equivalence between Riemann and Lebesgue integration.


\section{A Sliding-Window Algorithm for $\AvSTL$ Robustness}
\label{sec:algorithm}


We shall present an algorithm for computing $\AvSTL$ robustness. It
turns out that the presence of averaged modalities like
$\TDiaOp{I}$---with an apparent nonlocal nature---does not incur
severe computational overhead, at least for formulas
in which averaged modalities are not nested.
The algorithm is an adaptation of the one
in~\cite{DBLP:conf/cav/DonzeFM13} for $\STL$ robustness; the latter in turn relies on the 
\emph{sliding window minimum
algorithm}~\cite{DBLP:journals/njc/Lemire06}.
The algorithm's time complexity is linear with respect to the number of
timestamps in the input signal; it exhibits a practical speed, too, as
we will see later in~\S{}\ref{sec:experiments}.

Firstly we fix the class of signals to be considered.

\begin{mydefinition}[finitely piecewise-constant/piecewise-linear signal]
  A 1-dimensional signal $\sigma\colon \Rnn \to \R$ is 
  \emph{finitely piecewise-constant (FPC)}
  if it arises from a finite sequence 
  \begin{math}
    \bigl[\,   (t_{0},r_{0}),
      (t_{1},r_{1}),
      \dotsc,
      (t_{n},r_{n})
      \,\bigr]
  \end{math}
  of timestamped values, 
  via the correspondence
  \begin{math}
    \sigma(t) = r_{i} 
  \end{math}
  (for $t\in[t_{i}, t_{i+1})$).
    Here
    $0=t_{0}<\cdots<t_{n}$,
    $r_{i}\in\R$, and $t_{n+1}$ is deemed to be $\infty$.

    Similarly, a 1-dimensional signal 
    $\sigma\colon \Rnn \to \R$ is \emph{finitely piecewise-linear (FPL)}
    if it is identified with a finite sequence 
    \begin{math}
      \bigl[\,   (t_{0},r_{0}, q_{0}),
        \dotsc,
        (t_{n},r_{n},q_{n})
        \,\bigr]
    \end{math}
    of timestamped values, 
    via the correspondence
    \begin{math}
      \sigma(t)= r_{i} + q_{i}(t-t_{i}) 
    \end{math}
    (for $t\in[t_{i}, t_{i+1})$). Here $q_{i}\in R$ is the slope of $\sigma$
      in the interval $[t_{i}, t_{i+1})$.

 The definitions obviously extend to many-dimensional signals
$\sigma \colon \Rnn \to
 \R^\Var$.

We shall follow~\cite{DBLP:conf/formats/DonzeM10,DBLP:conf/cav/DonzeFM13} and  measure an algorithm's complexity in terms of
the number of timestamps ($n$ in the above); the latter is identified 
with the \emph{size} of a signal.
\end{mydefinition}




\begin{mydefinition}[robustness signal {$[\varphi]_{\sigma}$}]
  Let $\sigma : \Rnn \to \R^{\Var}$ be a signal, and $\varphi$ be an
 $\AvSTL$ formula.
 The \emph{positive robustness signal} of $\varphi$ over $\sigma$
 is the signal 
 $[\varphi]_{\sigma}^{+}\colon \Rnn\to\R$ defined by:
 $[\varphi]_{\sigma}^{+}(t)\Defeq\Robust{\sigma^{t}}{\varphi}^{+}$.
 Recall that $\sigma^{t}(t')=\sigma(t+t')$ is the $t$-shift of $\sigma$
 (Def.~\ref{def:signal}). The \emph{negative robustness signal}
 $[\varphi]_{\sigma}^{-}$ is defined in the same way.
\end{mydefinition}


\noindent
An averaged modality turns a piecewise-constant signal into a
piecewise-linear one.
\begin{mylemma}\label{lem:preservationOfPiecewiseConstLinear}
 \begin{enumerate}
  \item Let $\varphi$ be an averaging-free $\AvSTL$ formula.
	If a signal $\sigma$ is finitely piecewise-constant (or
	piecewise-linear), then so is $[\varphi]^{+}_{\sigma}$.
  \item Let $\varphi$ be an $\AvSTL$ formula without nested averaged
	modalities.
	If a signal $\sigma$ is finitely piecewise-constant, then
 $[\varphi]^{+}_{\sigma}$ is finitely piecewise-linear.
 \end{enumerate}
\noindent
The above holds for the negative robustness signal
 $[\varphi]^{-}_{\sigma}$, too.
\end{mylemma}
\begin{myproof}
 Straightforward by the induction on the construction of formulas.
 \myqed
\end{myproof}

Our algorithm for computing $\AvSTL$ robustness 
$\Robust{\sigma}{\varphi}$ will be focused on: 
1) a finitely piecewise-constant input signal $\sigma$; and 
2) an $\AvSTL$ formula $\varphi$ where averaged modalities are not nested.
In what follows, for presentation, we use the (non-averaged and averaged) 
eventually modalities $\DiaOp{I},\TDiaOp{I}$ in describing algorithms. 
Adaptation to other modalities is not hard; 
for complex formulas, we compute
the robustness signal $[\varphi]_{\sigma}$ by induction on $\varphi$.






\auxproof{
Unfortunately,
these propositions do not hold
for formulas that contain averaged modalities.
However,
we can still have the following.

\begin{myproposition}\label{prop:fpc2fpl}
  Let $\sigma : \Rnn \to \R$ and
  $\varphi_1, \varphi_2 \in \Fml$ be averaging-free formulas.
  If $[x]_{\sigma}$ is finitely piecewise-constant for any atomic formula $x$,
  then $[\varphi_1 \TUntil{[a,b]} \varphi_2]_{\sigma}$ is finitely piecewise-linear
  for any non-singular closed interval $[a,b]$.
\end{myproposition}

\begin{myproof}
  From Prop.~\ref{prop:fpc},
  $[\varphi_1]_{\sigma}$ and $[\varphi_2]_{\sigma}$ are both finitely piecewise constant.
  Let $(t_i)_{i \leq n}$ be the union set of those timestamps.
  We show that
  there exists finite partition of $R^+$
  such that
  $[\varphi_1 \TUntil{[a,b]} \varphi_2]_{\sigma}$ is linear
  in each.

  For any 
  $u \in \Rnn$,
  take any $u' \in \Rnn$ satisfying the both following conditions.
  \begin{itemize}
    \item $\{t_j \mid u \leq t_j \} = \{t_j \mid u' \leq t_j \}$
    \item $\{t_k \mid a+u \leq t_k \leq b+u \} = \{t_k \mid a+u'\leq t_k \leq b+u' \}$
  \end{itemize}
  Then,
  \[
    \begin{array}{ll}
      &\Robust{\sigma^{u'}}{\varphi_1 \TUntil{[a,b]} \varphi_2} 
      - \Robust{\sigma^u}{\varphi_1 \TUntil{[a,b]} \varphi_2}\\
      =&\Frac{1}{b-a} 
      \bigg( \Int_a^b \Robust{\sigma^{u'}}{\varphi_1 \UntilOp{[a,\tau]} \varphi_2} d\tau 
      - \Int_a^b \Robust{\sigma^{u}}{\varphi_1 \UntilOp{[a,\tau]} \varphi_2} d\tau \bigg)\\
    \end{array}
  \]
  and
  \begin{align*}
      & \Int_a^b \Robust{\sigma^{u'}}{\varphi_1 \UntilOp{[a,\tau]} \varphi_2} d\tau
      - \Int_a^b \Robust{\sigma^{u}}{\varphi_1 \UntilOp{[a,\tau]} \varphi_2} d\tau\\
      = &
      \Int_a^b 
      \Vee{\tau' \in [a, \tau] }
      \Big( \Robust{\sigma^{u'+\tau'}}{\varphi_2} \wedge \Wedge{\tau'' \in [0, \tau']} \Robust{\sigma^{u'+\tau''}}{\varphi_1} \Big) 
      d\tau \\
      &- \Int_a^b 
      \Vee{\tau' \in [a, \tau] }
      \Big( \Robust{\sigma^{u+\tau'}}{\varphi_2} \wedge \Wedge{\tau'' \in [0, \tau']} \Robust{\sigma^{u+\tau''}}{\varphi_1} \Big) 
      d\tau\\
      =&
      \Int_a^b 
      \Vee{\tau' \in [a, \tau] }
      \Big( \Robust{\sigma^{u'+\tau'}}{\varphi_2} \wedge \Wedge{\tau'' \in [0, \tau']} \Robust{\sigma^{u'+\tau''}}{\varphi_1} \Big) 
      d\tau \\
      &- \Int_a^b 
      \Vee{\tau' \in [a, \tau] }
      \Big( \Robust{\sigma^{u+\tau'}}{\varphi_2} \wedge \Wedge{\tau'' \in [0, \tau']} \Robust{\sigma^{u+\tau''}}{\varphi_1} \Big) 
      d\tau\\
      =&
      \Int_{a+u'}^{b+u'} 
      \Vee{\tau' \in [a+u', \tau] }
      \Big( \Robust{\sigma^{\tau'}}{\varphi_2} \wedge \Wedge{\tau'' \in [u', \tau']} \Robust{\sigma^{\tau''}}{\varphi_1} \Big) 
      d\tau \\
      &- \Int_{a+u}^{b+u} 
      \Vee{\tau' \in [a+u, \tau] }
      \Big( \Robust{\sigma^{\tau}}{\varphi_2} \wedge \Wedge{\tau'' \in [u, \tau']} \Robust{\sigma^{\tau''}}{\varphi_1} \Big) 
      d\tau\\
      =&
      \Int_{b+u}^{b+u'} 
      \Vee{\tau' \in [a+u', \tau] }
      \Big( \Robust{\sigma^{\tau'}}{\varphi_2} \wedge \Wedge{\tau'' \in [u', \tau']} \Robust{\sigma^{\tau''}}{\varphi_1} \Big) 
      d\tau \\
      &+\Int_{a+u'}^{b+u} 
      \Vee{\tau' \in [a+u', \tau] }
      \Big( \Robust{\sigma^{\tau'}}{\varphi_2} \wedge \Wedge{\tau'' \in [u', \tau']} \Robust{\sigma^{\tau''}}{\varphi_1} \Big) 
      d\tau \\
      &- \Int_{a+u'}^{b+u} 
      \Vee{\tau' \in [a+u, \tau] }
      \Big( \Robust{\sigma^{\tau}}{\varphi_2} \wedge \Wedge{\tau'' \in [u, \tau']} \Robust{\sigma^{\tau''}}{\varphi_1} \Big) 
      d\tau\\
      &- \Int_{a+u}^{a+u'} 
      \Vee{\tau' \in [a+u, \tau] }
      \Big( \Robust{\sigma^{\tau}}{\varphi_2} \wedge \Wedge{\tau'' \in [u, \tau']} \Robust{\sigma^{\tau''}}{\varphi_1} \Big) 
      d\tau.\\
    \end{align*}

  Here the second term and the third term will cancel, 
  otherwise some $\tau \in [a+u', b+u]$ exists such that 
  $\{ t_k \mid t_k \in [a+u, \tau]\} \neq \{ t_k \mid t_k \in [a+u', \tau]\}$
  or
  $\{ t_k \mid t_k \in [u, \tau]\} \neq \{ t_k \mid t_k \in [u', \tau]\}$,
  that conflicts with the conditions.
  Moreover,
  there exists no timestamps in $[b+u, b+u']$ and $[a+u, a+u']$,
  hence the integrands in the first term and the last term are
  constant functions in each domain of integration.
  Therefore,
  \[
    \Robust{\sigma^{u'}}{\varphi_1 \TUntil{[a,b]} \varphi_2} 
    - \Robust{\sigma^u}{\varphi_1 \TUntil{[a,b]} \varphi_2}
    = \Frac{C}{b-a} (u' - u)
  \]
  for some constant real number $C$ that does not depend on $u'$.
  Because $(t_i)_{i \leq n}$ is finite,
  the equivalence relation generated by the above conditions
  is also finite.
  Consequently, $[\varphi_1 \TUntil{[a,b]} \varphi_2]$ is finitely piecewise-linear.
  \myqed
\end{myproof}
}



\subsection{Donz\'e et al.'s Algorithm for $\STL$ Robustness}
\label{subsec:algoSTL}
We start with reviewing the algorithm~\cite{DBLP:conf/cav/DonzeFM13} for
$\STL$ robustness. 
Our algorithm for $\AvSTL$ robustness relies on it in two ways: 1)
the procedures for averaged modalities like $\TDiaOp{I}$ derive from those for non-averaged
modalities
in~\cite{DBLP:conf/cav/DonzeFM13}; and 2) we  use the algorithm
in~\cite{DBLP:conf/cav/DonzeFM13} itself for the non-averaged fragment of
$\AvSTL$. 

\begin{myremark}\label{rem:DonzeAlgoForPiecewiseLinear}
  The algorithm in~\cite{DBLP:conf/cav/DonzeFM13} computes the $\STL$
  robustness $\Robust{\sigma}{\varphi}$ for a finitely piecewise-\emph{linear}
  signal $\sigma$. 
  We need this feature e.g.\ for computing robustness of the formula
  \begin{math}
    \BoxOp{} (
    \mathtt{heavyBraking} 
  \end{math}
  \begin{math}
    \to \TDiaOp{[0,10]}\mathtt{airbag} )
  \end{math}: note that, by
  Lem.~\ref{lem:preservationOfPiecewiseConstLinear}, the robustness signal
  for
  $\TDiaOp{[0,10]}\mathtt{airbag} $ is piecewise-linear even if the input
  signal is piecewise-constant.
\end{myremark}


Consider  computing the robustness 
signal $[\DiaOp{[a,b]} \varphi]_{\sigma}$, assuming that the signal
$[\varphi]_{\sigma}$ is already given.\footnote{In the rest of~\S{}\ref{subsec:algoSTL}, for simplicity of
presentation, we assume that $[\varphi]_{\sigma}$ is
piecewise-constant. We note that the algorithm
in~\cite{DBLP:conf/cav/DonzeFM13} 
nevertheless extends to
piecewise-linear $[\varphi]_{\sigma}$.} The task calls for finding the supremum of
$[\varphi]_{\sigma}(\tau)$ over $\tau \in [t+a, t+b]$; and this must be done for each $t$. Naively doing so leads to quadratic complexity. 

\begin{figure}[tb]
\centering
\scalebox{.8}{
\begin{tabular}{ccccc}
$\cdots 
\stackrel[\text{bwd.}]{\text{slide}}{\longmapsto}$ 
&
\begin{tabular}{c}
 \includegraphics[width=.33\textwidth]{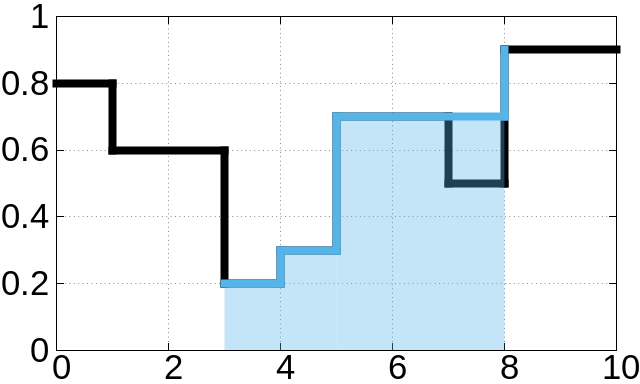}
\\
window in $[3,8]$
\end{tabular} 
& 
$
\stackrel[\text{bwd.}]{\text{slide}}{\longmapsto}$ 
& 
\begin{tabular}{c}
 \includegraphics[width=.33\textwidth]{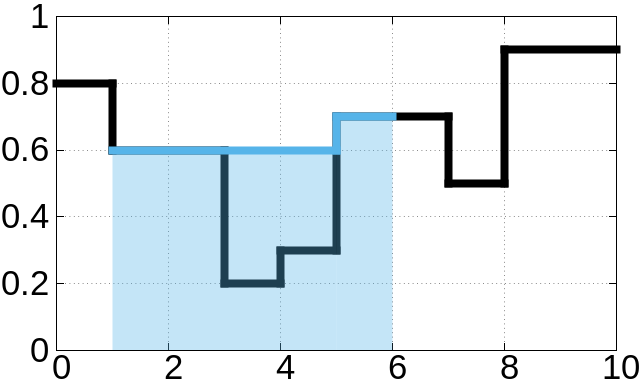}
\\
window in $[1,6]$
\end{tabular} 
 &
$
\stackrel[\text{bwd.}]{\text{slide}}{\longmapsto}\cdots $ 
\end{tabular}
}
\caption{A sliding window for computing
 $[\DiaOp{[0,5]} (x\ge
 0)]^+_{\sigma}$; the black line is the signal $\sigma$}
\label{fig:slidingWindow}
\scalebox{.8}{
  {\footnotesize
    \begin{math}
      \def\labelstyle{\textstyle}
      \def\twocellstyle{\textstyle}
      \xymatrix@R=1.5em@C+4em{
        {\begin{array}{c}
           \includegraphics[width=.3\textwidth]{pics/StQ/NewStQ3.png}
           \\
           \bigl[\;(3,0.2)\,(4,0.3)\,(5,0.7)\,(8,0.9)\;\bigr]
         \end{array} 
       }
       \ar@{|->}[r]^-{\text{slide}}_-{\text{backward}}
       \ar@{|->}[d]^-{\text{dequeue $(8,0.9)$}}
       &
       {\begin{array}{c}
          \includegraphics[width=.3\textwidth]{pics/StQ/NewStQ1.png}
          \\
          \bigl[\;(1,0.6)\,(5,0.7)\;\bigr]
        \end{array} 
      }
      \\
      {\begin{array}{c}
         \includegraphics[width=.3\textwidth]{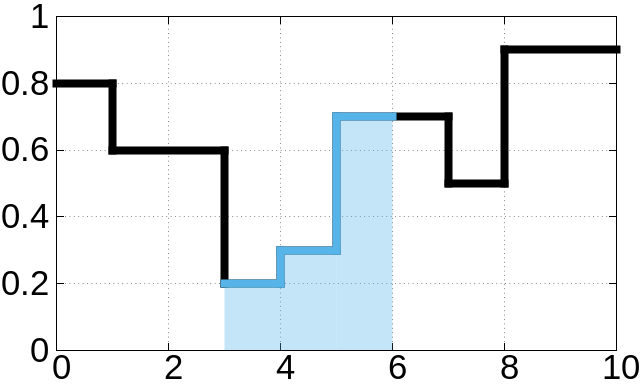}
         \\
         \bigl[\;(3,0.2)\,(4,0.3)\,(5,0.7)\;\bigr]
       \end{array} 
     }
     \ar@{|->}[r]^-{
       \begin{array}{r}
         \text{pop $(3,0.2)$}\quad\\
         \text{and $(4,0.3)$}
       \end{array}
     }
     &
     {\begin{array}{c}
        \includegraphics[width=.3\textwidth]{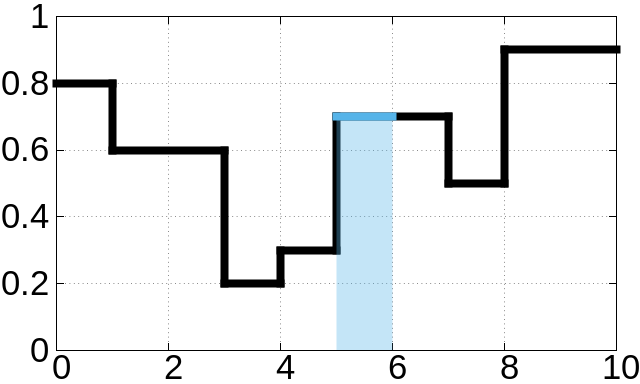}
        \\
        \bigl[\;(5,0.7)\;\bigr]
      \end{array} 
    }
    \ar@{|->}[u]^-{\text{push $(1,0.6)$}}
  }
\end{math}}}
\caption{Use of stackqueues and their operations, in the sliding window algorithm}
\label{fig:stackqueueForSlidingWindow}
\end{figure}

Instead Donz\'e et al.\ in~\cite{DBLP:conf/cav/DonzeFM13} employ a
\emph{sliding window} of size $b-a$ and let it scan the signal
$[\varphi]_{\sigma}$ from right to left. The scan happens once for all,
hence achieving linear complexity. See Fig.~\ref{fig:slidingWindow},
where we take 
$[\DiaOp{[0,5]} (x\ge 0)]^+_{\sigma}$ 
as an example, 
and the blue shaded area designates the position of the sliding window.
The window slides from $[3,8]$ to the closest position to the left where
its left-endpoint hits a new timestamped value of $[\varphi]_{\sigma}$, 
namely $[1,6]$.

\begin{wrapfigure}[3]{r}{0pt}
  \begin{tabular}{c}
    \includegraphics[clip,trim=0cm 2.5cm 0cm 2.5cm,width=.33\textwidth]{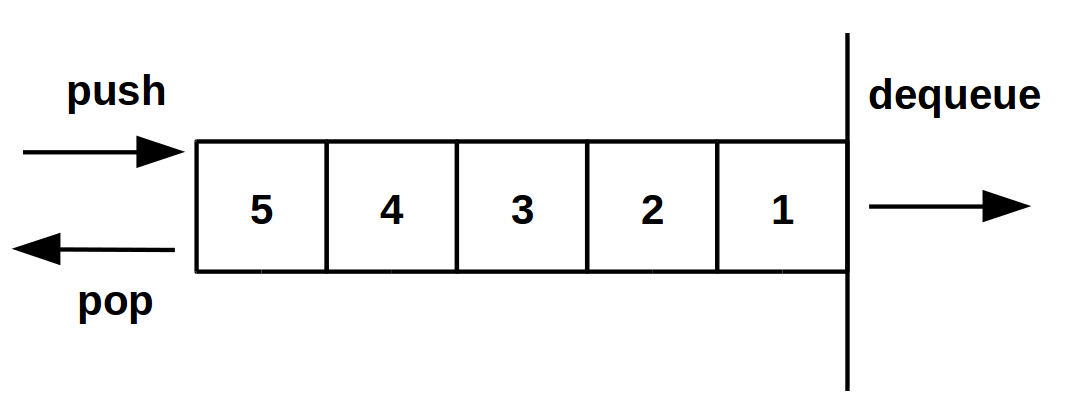}
    \\[-1em]
    a stackqueue
  \end{tabular} 
\end{wrapfigure}
It is enough to know the shape of the blue (partial) signal in
Fig.~\ref{fig:slidingWindow}, at each position of the window.
The blue signal denotes the (black) signal $\sigma$'s
local supremum within the window; more precisely, 
it denotes the value of the signal
$\Robust{\sigma^t}{\DiaOp{[0,\tau]}(x\ge 0)}^+$ at time $t+\tau$, where
$\tau \in [0,5]$ and $t$ is the leftmost position of the window.
We can immediately read off
 the signal
$[\DiaOp{[0,5]} (x\ge 0)]^+_{\sigma}$ 
from the blue signals: 
the former is the latter's value at the
 rightmost position of the window.

The keys in 
the algorithms
in~\cite{DBLP:conf/cav/DonzeFM13,DBLP:journals/njc/Lemire06}
lie in:
\begin{itemize}
 \item use of the \emph{stackqueue} data structure (depicted above on the right) for the
purpose of representing the blue (partial) signal in
Fig.~\ref{fig:slidingWindow}; and 
 \item use of the  operations \emph{push}, \emph{pop} and
       \emph{dequeue} for updating the blue signal.
\end{itemize}
See Fig.~\ref{fig:stackqueueForSlidingWindow}, where each entry of
a stackqueue is a timestamped value $(t,r)$. We see that the slide of the window, from top-left to
top-right in Fig.~\ref{fig:stackqueueForSlidingWindow},
is expressed by dequeue, pop and then push operations to stackqueues (in
Fig.~\ref{fig:stackqueueForSlidingWindow}: from top-left to bottom-left,
bottom-right and then top-right).
 Pseudocode for the algorithm is deferred to
Appendix~\ref{appendix:algoDia} due to lack of space.

\subsection{An Algorithm for $\AvSTL$ Robustness}
\label{subsec:algoAvSTL}
It turns out that the last algorithm is readily applicable to 
computing $\AvSTL$ robustness. Consider an averaged-eventually formula 
$\TDiaOp{[a,b]} \varphi$ as an example. What we have to compute is
the size of the shaded areas in Fig.~\ref{fig:slidingWindow} (see
also Fig.~\ref{fig:averagedDiamond}); and the shape of the blue 
signals in Fig.~\ref{fig:slidingWindow} carry just enough information
to do so.

Pseudocode for the adaptation of the previous algorithm 
(in~\S{}\ref{subsec:algoSTL}) to
$\TDiaOp{[a,b]} \varphi$ is found in
Algorithm~\ref{algo:tdia}.
Its complexity is linear  with respect to the number $n$ of the timestamp
values that represent the  signal $[\varphi]_{\sigma}$.

\begin{algorithm}[tbp]
  \caption{An algorithm for computing $[\TDiaOp{[a,b]} \varphi]_{\sigma}$}
  \label{algo:tdia}
  \begin{algorithmic}
    \Require An FPC signal $[\varphi]_{\sigma}$
    given as a sequence $(t_{0},r_{0}),\dotsc,(t_{n},r_{n})$ 
    \Ensure The FPL signal
    $[\TDiaOp{[a,b]} \varphi]_{\sigma}$
    \State $t_{\mathsf{temp}} := t_n - a$;
    \State $F := \bigl[ \;(t_{\mathsf{temp}}+a, [\varphi]_{\sigma}(t_{\mathsf{temp}} + a)) \;\bigr]$; 
    \Comment $F$ is the FPC signal $\tau \mapsto \Robust{\sigma^t}{\DiaOp{[a,\tau]}\varphi}$
    \State $s := (b-a) \cdot [\varphi]_{\sigma}(t_{\mathsf{temp}} + a)$;
    \Comment The area of $F$
    \State $G := \bigl[ \;(t_{\mathsf{temp}}, s / (b-a), 0) \;\bigr]$; 
    \Comment The FPC signal $[\TDiaOp{[a,b]} \varphi]_{\sigma}$
    \While{$t_{\mathsf{temp}} \geq 0$}
    \State $t_{\mathsf{old}} := t_{\mathsf{temp}}$;
    \State $t_{\mathsf{temp}} :=$ the greatest $t$ such that 
    $t < t_{\mathsf{old}} \wedge \bigl(\exists t_i.\, t + a = t_i \vee \exists (t', r') \in F. \; t+ b = t') \bigr)$;

    \State $\mathsf{Deq} := \{(t, r) \in F \mid t > t_{\mathsf{temp}} + b\}$; \quad $F := F \setminus \mathsf{Deq}$;
    \Comment{Dequeue old elements in $F$}
    \State $\mathsf{Pop} := \{(t, r) \in F \mid r \leq  [\varphi]_{\sigma}(t_{\mathsf{temp}} + a) \}$; \quad $F := F \setminus \mathsf{Pop}$;
    \Comment{Pop small elements in $F$}
    \State $t_{\mathsf{Pop}} := \mathsf{min} \{t \mid(t,r) \in F \; \text{ or } \; t = t_{\mathsf{temp}} + b\}$; 
    \State $F := \big[ (t_{\mathsf{temp}}+a,  [\varphi]_{\sigma}(t_{\mathsf{temp}} + a))\big] \cup F$ 
    \Comment{Push the left endpoint of the window to $F$}
    \State $r_{\mathsf{left}} := \Min\{ r \mid (t,r) \in F \}$;
    \State $r_{\mathsf{right}} := \Max\{ r \mid (t,r) \in F \}$;
    \State $s := s
    - (t_{\mathsf{old}} - t_{\mathsf{temp}}) \cdot r_{\mathsf{right}} 
    - \mathsf{area}(\mathsf{Pop}) 
    + (t_{\mathsf{Pop}} - (t_{\mathsf{temp}} +a )) \cdot r_{\mathsf{left}}$
    \State $G := \{(t_{\mathsf{temp}}, s/(b-a),r_{\mathsf{right}} - r_{\mathsf{left}})\} \cup G$
    \EndWhile
  \end{algorithmic}
\end{algorithm}

An algorithm for the averaged-henceforth formula
$[\TBoxOp{[a,b]} \varphi]_{\sigma}$ is similar. Extensions to
averaged-until and averaged-release operators are possible, too; they
use doubly-linked lists in place of stackqueues (see Appendix~\ref{appendix:algoTUntil}).
Combining with the algorithm in~\S{}\ref{subsec:algoSTL} to deal with
non-averaged temporal operators, we have the following 
complexity result. The complexity is the same as for $\STL$~\cite{DBLP:conf/cav/DonzeFM13}.



\begin{mytheorem}\label{thm:complexity_rel}
  Let $\varphi$ be an $\AvSTL$ formula 
  in which averaged modalities are not nested.
  Let $\sigma$ be a finitely piecewise-constant signal.
  Then there exists an algorithm
  to compute
  $\Robust{\sigma}{\varphi}^{+}$
  with time-complexity in 
  $\mathcal{O}(d^{|\varphi|}  |\varphi| |\sigma|)$
  for some constant $d$.

  The same is true for the negative robustness
  $\Robust{\sigma}{\varphi}^{-}$.
  \myqed
\end{mytheorem}

\begin{myremark}
 The reason for our restriction to finitely piecewise-constant input
 signals is hinted in Rem.~\ref{rem:DonzeAlgoForPiecewiseLinear}; let us
 further elaborate on it.
 There the averaged modality $\TDiaOp{[0,10]}$ turns a piecewise-constant signal into a
 piecewise-linear one
 (Lem.~\ref{lem:preservationOfPiecewiseConstLinear}); and then the
 additional Boolean connectives and non-averaged
 modalities (outside  $\TDiaOp{[0,10]}$) are taken care of by the algorithm
 in~\cite{DBLP:conf/cav/DonzeFM13}, one that is restricted
 to piecewise-linear input.

 It is not methodologically hard to extend this workflow to
 piecewise-\emph{polynomial} input signals (hence to nested averaged
 modalities as well). Such an extension however calls
 for computing local suprema of polynomials, as well as their
 intersections---tasks that are drastically easier with affine
 functions. We therefore expect  the extension to
 piecewise-polynomial signals to be computationally
 much more expensive.
\end{myremark}

\section{Enhanced Falsification: Implementation and Experiments}
\label{sec:experiments}

\begin{table}[ptb]
  \scriptsize
  \centering
  \begin{minipage}{\textwidth}
    \textbf{Problem 1. } Falsification means finding an input signal 
    that keeps the engine speed $\omega$ below 2000 rpm, for $T$
    seconds. The bigger $T$ is, the harder the problem is.
    We applied $\Diamond$-refinement.
  \end{minipage}
  \begin{tabular}{c||r|r|r|r|r|r|r|r|r}
    \textbf{Problem 1}
    &\multicolumn{3}{|c|}{$T = 20$} &\multicolumn{3}{|c|}{$T = 30$} &\multicolumn{3}{|c}{$T = 40$}\\ \hline
    Specification & \Succ & Iter. & Time & \Succ & Iter. & Time &
    \Succ & Iter. & Time \\
    to be falsified
    & $\mathbf{/100}$ & (\Succ) & (\Succ) &$\mathbf{/100}$ & (\Succ) & (\Succ)&$\mathbf{/100}$  & (\Succ) & (\Succ)\\  \hline\hline
    $\DiaOp{[0,T]}{(\omega \geq 2000)}$ 
    & 100& 128.8& 20.2& 81& 440.9& 82.5& 32& 834.3& 162.9\\
    &    & 128.8& 20.2&   & 309.7& 59.0&   & 482.2&  94.4\\\hline
    $\TDiaOp{[0,T]}{(\omega \geq 2000)}$
    & 100& 123.9& 22.9& 98& 249.8& 46.1 & 81& 539.6& 110.9\\
    &    & 123.9& 22.9&   & 234.5& 43.4 &   & 431.6&  89.2\\
  \end{tabular}

  \vspace{1em}
  \begin{minipage}{\textwidth}
    \textbf{Problem 2.} Falsification means finding an input signal 
    that keeps $\omega$ within a range of 3500--4500 rpm for $T$
    consecutive seconds, at a certain stage. 
    We applied $\Diamond$-refinement.
  \end{minipage}
  \begin{tabular}{c||r|r|r}
    \textbf{Problem 2}
    &\multicolumn{3}{|c}{$T = 10$} \\ \hline
    Specification & \Succ & Iter. & Time\\
    to be falsified
    & $\mathbf{/100}$ & (\Succ) & (\Succ) \\  \hline\hline
    $\BoxOp{}\DiaOp{[0, T]}(\omega \leq 3500 \vee \omega \geq 4500)$
    & 45& 625.4& 209.1\\
    &   & 167.7&  56.1\\\hline
    $\BoxOp{}\TDiaOp{[0, T]}(\omega \leq 3500 \vee \omega \geq 4500)$
    & 74& 442.0&  154.3\\
    &   & 245.9&   86.6\\
  \end{tabular}

  \vspace{1em}
  \begin{minipage}{\textwidth}
    \textbf{Problem 3.} Falsification means finding an input signal 
    that shifts the gear into the fourth within $T$ seconds. 
    The smaller $T$ is, the harder the problem is. Here $\mathtt{gear}_4$
    is a propositional variable.
    We applied $\Box$-refinement.
  \end{minipage}
  \begin{tabular}{c||r|r|r|r|r|r|r|r|r}
    \textbf{Problem 3}
    &\multicolumn{3}{|c|}{$T = 4$} &\multicolumn{3}{|c|}{$T = 4.5$} &\multicolumn{3}{|c}{$T = 5$}\\ \hline
    Specification & \Succ & Iter. & Time & \Succ & Iter. & Time &
    \Succ & Iter. & Time \\
    to be falsified
    & $\mathbf{/20}$ & (\Succ) & (\Succ) &$\mathbf{/20}$ & (\Succ) & (\Succ)&$\mathbf{/20}$  & (\Succ) & (\Succ)\\  \hline\hline
    $\BoxOp{[0,T]}{\neg \mathtt{gear}_4}$ 
    & 0& 1000& 166.9& 11& 742.8 & 122.9& 18& 449.0 &  71.8 \\ 
    &  &    --&     --&   & 532.3 &  87.5&   & 387.7 &  61.9 \\ \hline
    $\BoxOp{[0,T]}{\neg \mathtt{gear}_4}$ 
    & 17& 570.1& 94.0& 20& 250.5& 40.3& 20& 107.5& 17.6\\
    $\wedge \TBoxOp{[T,10]}{\neg \mathtt{gear}_4}$
    &   & 494.2& 81.8&   & 250.5& 40.3&   & 107.5& 17.6\\
  \end{tabular}

  \vspace{1em}
  \begin{minipage}{\textwidth}
    \textbf{Problem 4.} Falsification means finding input  
    with which the gear never stays in the third consecutively for $T$ seconds.
    The smaller $T$ is, the harder the problem is.
    Here $\mathtt{gear}_3$
    is a propositional variable.
    We applied $\Box$-refinement.
  \end{minipage}
  \begin{tabular}{c||r|r|r|r|r|r}
    \textbf{Problem 4}
    &\multicolumn{3}{|c|}{$T = 1$} &\multicolumn{3}{|c}{$T = 2$} \\ \hline
    Specification & \Succ & Iter. & Time & \Succ & Iter. & Time \\
    to be falsified
    & $\mathbf{/20}$ & (\Succ) & (\Succ) &$\mathbf{/20}$ & (\Succ) & (\Succ)\\  \hline\hline
    $\DiaOp{}\big(\BoxOp{[0, T]}\mathtt{gear}_3\big)$
    & 14& 556.1& 132.0& 20& 82.8& 20.6\\
    &   & 365.8&  87.1&   & 82.8& 20.6\\\hline
    $\DiaOp{}\big(\BoxOp{[0, T]}\mathtt{gear}_3 \wedge \TBoxOp{[T, 10]}\mathtt{gear}_3 \big)$
    & 20& 105.1&  36.3& 20& 29.7& 10.2\\
    &   & 105.1&  36.3& 20& 29.7& 10.2\\
  \end{tabular}

  \vspace{1em}
  \begin{minipage}{\textwidth}
    \textbf{Problem 5.} Falsification means finding input
    that violates the following requirement: \emph{after the gear is
      shifted, it stays the same for $T$ seconds.}
    (the smaller $T$, the harder).
    $\mathtt{gear}_{1},\dotsc,\mathtt{gear}_{4}$ are propositional variables.
    We applied $\Box$-refinement.
  \end{minipage}
  \begin{tabular}{c||r|r|r|r|r|r|r|r|r}
    \textbf{Problem 5} ($\varepsilon = 0.04$)
    &\multicolumn{3}{|c|}{$T = 0.8$} &\multicolumn{3}{|c|}{$T = 1$} &\multicolumn{3}{|c}{$T = 2$}\\ \hline
    Specification & \Succ & Iter. & Time & \Succ & Iter. & Time &
    \Succ & Iter. & Time \\
    to be falsified
    & $\mathbf{/20}$ & (\Succ) & (\Succ) & $\mathbf{/20}$& (\Succ) & (\Succ)& $\mathbf{/20}$ & (\Succ) & (\Succ)\\  \hline\hline
    {\scriptsize $\bigwedge_{i=1,\dotsc,4}\BoxOp{}\Big(\big( \neg \mathtt{gear}_i \wedge \DiaOp{[0,\varepsilon]}\mathtt{gear}_i \big)$}
    & 2& 972.5 & 402.5& 19& 356.8& 155.6& 20& 27.4& 11.8\\
    {\scriptsize $\to \big(\BoxOp{[\varepsilon, T+\varepsilon]} \mathtt{gear}_i \big) \Big)$}
    &  & 724.5 & 297.8&   & 322.9& 140.9&   & 27.4& 11.8\\ \hline
    {\scriptsize $\bigwedge_{i=1,\dotsc,4}\BoxOp{}\Big(\big(\neg \mathtt{gear}_i \wedge \DiaOp{[0,\varepsilon]}\mathtt{gear}_i \big)$}
    & 12& 561.1& 349.1& 20& 93.1& 57.8& 20& 42.7& 26.9\\
    {\scriptsize $\to \big(\BoxOp{[\varepsilon, T+\varepsilon]} \mathtt{gear}_i\wedge \TBoxOp{[T+\varepsilon, 5]} \mathtt{gear}_i\big)\Big)$}
    &   & 268.5& 167.3&   & 93.1& 57.8&   & 42.7& 26.9\\
  \end{tabular}

  \vspace{1em}
  \begin{minipage}{\textwidth}
    \textbf{Problem 6.} 
    Falsification means finding an input signal 
    that steers the vehicle speed $v$ over 85 kph within $T$ seconds,
    while keeping the engine speed $\omega$ below 4500 rpm.
    The smaller $T$ is, the harder the problem is.
    We applied $\Box$-refinement.
  \end{minipage}
  \begin{tabular}{c||r|r|r|r|r|r}
    \textbf{Problem 6}
    &\multicolumn{3}{|c|}{$T = 10$} &\multicolumn{3}{|c}{$T = 12$} \\ \hline
    Specification & \Succ & Iter. & Time & \Succ & Iter. & Time \\
    to be falsified
    & $\mathbf{/20}$ & (\Succ) & (\Succ) &$\mathbf{/20}$ & (\Succ) & (\Succ)\\  \hline\hline
    $\BoxOp{[0, T]}{(v \leq 85)} \vee \DiaOp{}{(\omega \geq 4500)}$
    & 12& 714.9& 141.4& 17& 374.5& 72.2\\
    &   & 524.9& 108.1&   & 264.1& 51.2\\\hline
    $\big(\BoxOp{[0, T]}{(v \leq 85)} \wedge \TBoxOp{[T, 20]}{(v \leq 85)} \big)$
    &12& 766.7& 149.0& 20& 423.6& 85.7\\
    $\vee \DiaOp{}{(\omega \geq 4500)}$
    &  & 611.2& 118.9&   & 423.6& 85.7\\
  \end{tabular}
  \caption{Experiment results. Time is in seconds. The ``Succ.''
    columns show how many trials  succeeded among the designated number of trials;
    the ``Iter.'' columns show the average number of iterations of the
    S-TaLiRo loop, executed in each trial (max.\ 1000); and the ``Time'' columns show
    the average time that each trial took. For the last two we also show
    the average over \emph{successful} trials.}
  \label{table:result}
\end{table}

\begin{wrapfigure}[13]{r}{.338\textwidth}
  \includegraphics[clip,trim=0cm 15.5cm 25cm 0cm,width=.338\textwidth]
  {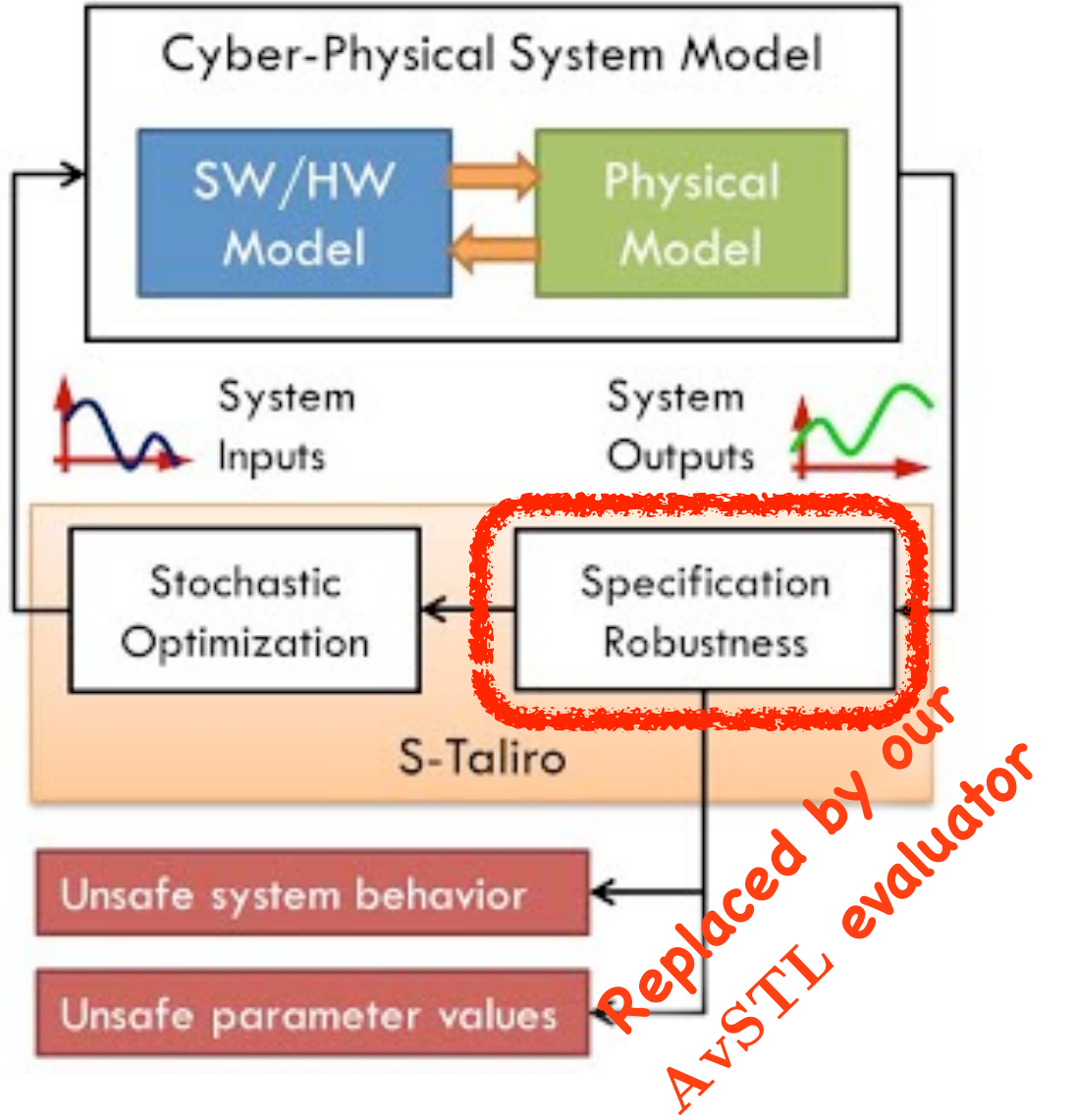}
  \vspace*{-2em}
  \caption{An overview of S-TaLiRo (from~\cite{WEB:S_TaLiRo}), with our modification}
  \label{fig:staliro}
\end{wrapfigure}
We claim that our logic $\AvSTL$ achieves a good balance between
expressivity---that communicates a designer's intention more faithfully
to a falsification
solver---and computational cost, thus contributing to the model-based
development of cyber-physical systems. In this section we present our
implementation that combines: 1)
S-TaLiRo~\cite{DBLP:conf/tacas/AnnpureddyLFS11}, one of the
state-of-art falsification solvers that relies on robust MTL semantics and
stochastic optimization;  and 2) the \emph{$\AvSTL$ evaluator}, 
an implementation of the algorithm
in~\S{}\ref{subsec:algoAvSTL}. Our experiments are on automotive
examples of falsification problems;
the results indicate that (refinement of specifications by) $\AvSTL$ brings considerable performance
improvement.

\subsubsection{Implementation}
S-TaLiRo~\cite{DBLP:conf/tacas/AnnpureddyLFS11} is 
``a Matlab toolbox that searches for trajectories of minimal robustness
in Simulink/Stateflow''~\cite{WEB:S_TaLiRo}. Recall the formalization of 
a falsification problem (\S{}\ref{sec:introduction}). S-TaLiRo's input
is:  1) a
model $\mathcal{M}$ that is a Simulink/Stateflow model; and 2) a specification
$\varphi$ that is an $\STL$ formula. 

S-TaLiRo employs stochastic simulation in the following 
\emph{S-TaLiRo loop}: 
\begin{enumerate}
 \item Choose an input signal $\sigma_{\mathsf{in}}$  randomly.
 \item Compute the output signal
       $\mathcal{M}(\sigma_{\mathsf{in}})$ with Simulink.
 \item Compute the robustness
  $\Robust{\mathcal{M}(\sigma_{\mathsf{in}})}{\varphi}$.
 \item If the robustness is $\leq 0$ then return $\sigma_{\mathsf{in}}$
  as a critical path. Otherwise choose a new $\sigma_{\mathsf{in}}$ 
  (hopefully with a smaller robustness)
    and go back to Step 2.
\end{enumerate}
Our  modification of S-TaLiRo consists of: 1) changing the specification formalism from
$\STL$ to $\AvSTL$ (with the hope that the robustness
$\Robust{\mathcal{M}(\sigma_{\mathsf{in}})}{\varphi}^+$ carries more
information to be exploited in stochastic optimization); and 2) using, in
 Step 3 of the above loop, the $\AvSTL$
evaluator based on the sliding-window algorithm
in~\S{}\ref{sec:algorithm}. 
See  Fig.~\ref{fig:staliro}.

\subsubsection{Experiments}
As a model $\mathcal{M}$ we used the automatic transmission model
from~\cite{HoxhaAF14arch1}, where it is offered ``as benchmarks
        for testing-based falsification''~\cite{HoxhaAF14arch1}.
The same model has been used 
in several works~\cite{DBLP:conf/pts/YangHF12, 6315384, DBLP:conf/hybrid/JinDDS13}.
The model has two input ports (\emph{$\mathtt{throttle}$} and
\emph{$\mathtt{brake}$}) and six output ports (the engine speed $\omega$, 
the vehicle speed $v$, and four mutually-exclusive Boolean ports
 $\mathtt{gear}_{1},\dotsc,\mathtt{gear}_{4}$ for the current gear).
Further illustration  is in Appendix~\ref{appendix:ATModel}.
%
%
As a specification $\varphi$ to  falsify, we took six examples from~\cite{HoxhaAF14arch1},
sometimes with minor modifications. They constitute Problems 1--6 in Table~\ref{table:result}.

Our goal is to examine the effect of our modification to S-TaLiRo.
For the  model $\mathcal{M}$ (that is fixed) and each of the six specifications
$\varphi$, experiments are done with:
\begin{itemize}
 \item $\mathcal{M}$ and the original $\STL$ formula $\varphi$, as a
       control experiment;
       and
 \item $\mathcal{M}$ and the $\AvSTL$ formula $\varphi'$ that is
       obtained from $\varphi$ as a
       refinement. The latter specifically involves \emph{$\Diamond$-refinement} and
       \emph{$\Box$-refinement} described in~\S{}\ref{subsec:soundnessOfEnrichments}.
\end{itemize}
Faster, or more frequent, falsification in the latter setting witnesses
effectiveness of our $\AvSTL$ approach. We note that
falsifying $\varphi'$ indeed means falsifying $\varphi$, because of the
soundness of the refinement (Prop.~\ref{prop:diaRefinement} and~\ref{prop:BoxRefinement}).

A single falsification \emph{trial} consists of at most 1000 \emph{iterations} of the S-TaLiRo
loop. For each specification $\varphi$ (i.e. for each problem in
Table~\ref{table:result}) we made 
20--100 falsification trials, 
sometimes
with different parameter values $T$. We made multiple trials
because of the stochastic nature of S-TaLiRo.

\subsubsection{Experiment Results and Discussion}
The experiment results are in Table~\ref{table:result}.
We used
Matlab R2014b and
S-TaLiRo ver.1.6 beta
on
ThinkPad T530 with Intel Core i7-3520M 2.90GHz CPU with 3.7GB memory.
The OS is Ubuntu14.04 LTS (64-bit).


 Notable performance improvement is observed in Problems 3--5, 
especially in their harder instances. For
example, our $\AvSTL$ enrichment made 17 out of 20 trials succeed in
Problem 3 ($T=4$), while no trials succeeded with the original $\STL$
specification. A similar extreme performance gap is observed also in
Problem 5 ($T=0.8$).

Such performance improvement in Problems 3--5  is not surprising. The
specifications for these problems
are concerned solely with the propositional variables
$\mathtt{gear}_{i}$ (cf.\ Rem.~\ref{rem:propVar});
and the space
robustness semantics for $\STL$ assigns to these specifications only $0$
or $1$
(but no values in-between) as their
truth 
values. We can imagine such ``discrete'' robustness values give few clues
to stochastic optimization algorithms.

Both of 
 $\Diamond$- and
       $\Box$-refinement
 in~\S{}\ref{subsec:soundnessOfEnrichments}
turn out to be helpful. The latter's effectiveness is observed in
Problems 3--5; the former improves a success rate from 32/100 to 81/100
in Problem 1 ($T=40$). 

Overall, the experiment results seem to support our claim that the complexity of
(computing robustness values in) $\AvSTL$ is tractable. There is no big
difference in the time 
each iteration takes, between the $\STL$ case and the $\AvSTL$ case.

\section{Conclusions and Future Work}
We introduced $\AvSTL$, an extension of $\STL$ with \emph{averaged}
temporal
operators. It adequately captures both space and time robustness; and
we presented an algorithm for computing  robustness that is 
linear-time with respect to the ``size'' of an input signal. Its use
in falsification of CPS is demonstrated by our prototype that modifies S-TaLiRo.

As future work, we wish to compare our averaged temporal operators with
other quantitative temporal operators, among which are the
\emph{discounting}
ones~\cite{DBLP:dblp_journals/jacm/AlurFH96,DBLP:conf/tacas/AlmagorBK14}. The
latter are closely related to \emph{mean-payoff}
conditions~\cite{Ehrenfeucht1979, DBLP:conf/lics/ChatterjeeHJ05} as well
as to \emph{energy
constraints}~\cite{DBLP:conf/formats/BouyerFLMS08,DBLP:conf/hybrid/BrenguierCR14},
all of which are studied principally in the context of automata theory.

Application of $\AvSTL$ to problems other than falsification is another
important direction. Among them is \emph{parameter synthesis}, another
task that S-TaLiRo is capable of.
We are now looking at application to
\emph{sequence classification} (see
e.g.~\cite{DBLP:conf/hybrid/KongJAGB14}), too,
whose significant role in model-based development of CPS is widely acknowledged.

\bibliographystyle{plain} 
\bibliography{./RelSTL} %

\newpage
\appendix
\section{Algorithms}
\subsection{An $\STL$ Algorithm for Computing
  $[\DiaOp{[a,b]} \varphi]_{\sigma}$, from~\cite{DBLP:conf/cav/DonzeFM13}  }
\label{appendix:algoDia}
In Algorithm~\ref{algo:dia} is pseudocode for 
computing the signal $[\DiaOp{[a,b]}\varphi]_{\sigma}$,
 given the signal $[\varphi]_{\sigma}$.
Its intuitions are found
in~\S{}\ref{subsec:algoSTL}.


\begin{algorithm}
  \caption{An algorithm for computing $[\DiaOp{[a,b]} \varphi]_{\sigma}$}
  \label{algo:dia}
    \begin{algorithmic}
      \Require A FPC signal $[\varphi]_{\sigma}$ given as a sequence 
      $(t_{0},r_{0}),\dotsc,(t_{n},r_{n})$ 
      \Ensure The FPC signal $[\DiaOp{[a,b]} \varphi]_{\sigma}$
      \State $t_{\mathsf{temp}} := t_n - a$; 
      \State $F := \bigl[ \;(t_{\mathsf{temp}}+a, [\varphi]_{\sigma}(t_{\mathsf{temp}} + a)) \;\bigr]$; 
      \Comment $F$ is the FPC signal 
      $\tau \mapsto \Robust{\sigma^{t_{\mathsf{temp}}}}{\DiaOp{[a,\tau]}\varphi}$
      \State 
      $G := \bigl[ \;(t_{\mathsf{temp}}, [\varphi]_{\sigma}(t_{\mathsf{temp}} + a)) \;\bigr]$; 
      \Comment $G$ is the FPC signal $[\DiaOp{[a,b]} \varphi]_{\sigma}$
      \While{$t_{\mathsf{temp}} \geq 0$}
      \State $t_{\mathsf{temp}} :=$ the greatest $t$ such that 
      $t < t_{\mathsf{temp}} \wedge \bigl(\exists t_i.\, t + a = t_i \vee \exists (t', r') \in F. \; t+ b = t') \bigr)$;
      \State $F := F \setminus \{(t, r) \mid t > t_{\mathsf{temp}} + b\}$; \Comment{Dequeue old elements in $F$}
      \State $F := F \setminus \{(t, r) \mid r \leq [\varphi]_{\sigma}(t_{\mathsf{temp}} + a) \}$; 
      \Comment{Pop elements in $F$ that are too small}
      \State $F := \big[ \;(t_{\mathsf{temp}}+a, [\varphi]_{\sigma}(t_{\mathsf{temp}} + a) )\;\big] \cup F$;
      \Comment{Push the left endpoint of the window to $F$}
      \State $r_{\mathsf{right}} := \Max \{ r \mid (t,r)\in F \}$;
      \State $G := \{(t_{\mathsf{temp}}, r_{\mathsf{right}})\} \cup G$;
      \Comment{Add a  timestamped value}
      \EndWhile
    \end{algorithmic}
\end{algorithm}

\subsection{An Algorithm for Computing $[\varphi_1 \TUntil{[a,b]} \varphi_2]_{\sigma}$ }
\label{appendix:algoTUntil}

 Algorithm~\ref{algo:tdia}
is an algorithm for computing $[\TDiaOp{[a,b]}
\varphi]_{\sigma}$
that is linear-time
with respect to the ``size'' of $[\varphi]_{\sigma}$.
We can compute $[\varphi_1 \TUntil{[a,b]} \varphi_2]_{\sigma}$
in linear-time, similarly,
by employing a sliding-window 
that stands for a piecewise constant function
\[
  \begin{array}{cccc}
    F:\quad & [a, b] & \;\longrightarrow\; & \R \\
       & \tau & \longmapsto & \Robust{\sigma^t}{\varphi_1 \UntilOp{[a, \tau]} \varphi_2}\enspace.
  \end{array}
\]
The sliding of the window corresponds to the change of the value of $t$.
For efficient implementation of such sliding we rely on the following
proposition. It is derived essentially from 
the equivalence $\varphi_1 \UntilOp{[a, \tau]} \varphi_2 \cong 
\DiaOp{[a, \tau]} \varphi_2 \wedge \BoxOp{[0,a]} \varphi_1 \UntilOp{}
\varphi_2$, an equivalence also used  in~\cite{DBLP:conf/cav/DonzeFM13}.
\begin{myproposition}\label{prop:TUntilAlgo}
Assume
that the signal
 $[\BoxOp{[0,a]}\varphi_{1}\UntilOp{} \varphi_{2}]_{\sigma}$ is constant
 in the interval $[0,\delta)$. Then we have
\begin{equation}\label{eq:propTUntilAlgo}
     \Robust{\sigma}{\varphi_1 \UntilOp{[a, \tau + \delta]} \varphi_2}
    = 
    \bigl(\,\Robust{\sigma^\delta}{\varphi_1 \UntilOp{[a, \tau]} \varphi_2} 
    \sqcup \Robust{\sigma}{\DiaOp{[a,a+\delta]} \varphi_2}\,\bigr)
    \sqcap \Robust{\sigma}{\BoxOp{[0,a]} \varphi_1 \UntilOp{} \varphi_2}\enspace.
\end{equation}
\end{myproposition}
\begin{myproof}
  \begin{align*} 
&\textrm{(RHS)} 
\\
    &=  \bigl(\Robust{\sigma^\delta}{\varphi_1 \UntilOp{[a, \tau]} \varphi_2} 
      \sqcap \Robust{\sigma}{\BoxOp{[0,a]} \varphi_1 \UntilOp{} \varphi_2}\bigr)
      \sqcup
      \bigl(\Robust{\sigma}{\DiaOp{[a, a+\delta]}{\varphi_2}}
      \sqcap \Robust{\sigma}{\BoxOp{[0,a]} \varphi_1 \UntilOp{}
   \varphi_2}\bigr)
\\
   &\qquad\qquad\qquad\qquad
    \text{by distributing $\sqcap$ over $\sqcup$}
\\
      &=  \bigl(\Robust{\sigma^\delta}{\varphi_1 \UntilOp{[a, \tau]} \varphi_2} 
      \sqcap \Robust{\sigma}{\BoxOp{[0,a]} \varphi_1 \UntilOp{} \varphi_2}\bigr)
      \sqcup \Robust{\sigma}{\varphi_1 \UntilOp{[a, a+\delta]}
   \varphi_2}
\\
   &\qquad\qquad\qquad\qquad
    \text{by the above equivalence $\varphi_1 \UntilOp{[a, \tau]} \varphi_2 \cong 
\DiaOp{[a, \tau]} \varphi_2 \wedge \BoxOp{[0,a]} \varphi_1 \UntilOp{}
\varphi_2$} 
\\
       &= 
          \big((
          \Robust{\sigma^\delta}{\DiaOp{[a, \tau]} \varphi_2} 
          \sqcap \Robust{\sigma^\delta}{\BoxOp{[0,a]} \varphi_1 \UntilOp{} \varphi_2})
          \sqcap \Robust{\sigma}{\BoxOp{[0,a]} \varphi_1 \UntilOp{} \varphi_2} \big)
          \sqcup \Robust{\sigma}{\varphi_1 \UntilOp{[a, a+\delta]} \varphi_2}
\\
   &\qquad\qquad\qquad\qquad
    \text{by the same equivalence} 
\\
       &= 
          \big((
          \Robust{\sigma}{\DiaOp{[a+\delta, \tau+\delta]} \varphi_2} 
          \sqcap \Robust{\sigma}{\BoxOp{[0+\delta,a+\delta]} \varphi_1 \UntilOp{} \varphi_2})
          \sqcap \Robust{\sigma}{\BoxOp{[0,a]} \varphi_1 \UntilOp{} \varphi_2} \big)
          \sqcup \Robust{\sigma}{\varphi_1 \UntilOp{[a, a+\delta]} \varphi_2}
\\
   &= (
        \Robust{\sigma}{\DiaOp{[a+\delta, \tau+\delta]} \varphi_2} 
        \sqcap \Robust{\sigma}{\BoxOp{[0,a+\delta]} \varphi_1 \UntilOp{} \varphi_2})
        \sqcup \Robust{\sigma}{\varphi_1 \UntilOp{[a, a+\delta]}
   \varphi_2}
\\
   &\qquad\qquad\qquad\qquad
    \text{by the assumption that $[\BoxOp{[0,a]}\varphi_{1}\UntilOp{} \varphi_{2}]_{\sigma}$ is constant
 in  $[0,\delta)$} 
\\
     &=
        \Robust{\sigma}{\varphi_1 \UntilOp{[a+\delta, \tau+\delta]} \varphi_2}
        \sqcup \Robust{\sigma}{\varphi_1 \UntilOp{[a, a+\delta]}
   \varphi_2}\qquad
    \text{again by the same equivalence} 
\\
     &= (\mathrm{LHS})\enspace. \tag*{\myqed}
  \end{align*}
\end{myproof}

Roughly speaking, the equality~(\ref{eq:propTUntilAlgo}) shows how the
signal \emph{after} sliding ($\Robust{\sigma}{\varphi_1 \UntilOp{[a,
\tau + \delta]} \varphi_2}$ on the left-hand side) can be computed from
the signal \emph{before} sliding (the first term 
$\Robust{\sigma^\delta}{\varphi_1 \UntilOp{[a, \tau]} \varphi_2}$
on the right-hand side). 

In Algorithm~\ref{algo:tuntil} pseudocode is found for 
computing 
$[\varphi_1 \TUntil{[a,b]} \varphi_2]_{\sigma}$. Compared to
Algorithm~\ref{algo:tdia} a principal addition is \emph{truncation} of
big elements ($\mathsf{Trunc}$ in Algorithm~\ref{algo:tuntil}; it
corresponds
to taking $\sqcap$ in~(\ref{eq:propTUntilAlgo})). To realize such a
truncation operation
efficiently, we use a \emph{doubly-linked list} as a data
structure---in place of a stackqueue---so that it allows push and pop
from each side.

It is not hard to see that the time-complexity of
Algorithm~\ref{algo:tuntil} is linear in $n+m$. Note also that the signal
$[\BoxOp{[0,a]} \varphi_1 \UntilOp{} \varphi_2]_{\sigma}$ 
(input to Algorithm~\ref{algo:tuntil}) can be computed
efficiently, from  
the signals $[\varphi_1]_{\sigma}$ and $[\varphi_2]_{\sigma}$,
thanks to the algorithm presented in~\cite{DBLP:conf/cav/DonzeFM13}.
\begin{algorithm}
  \caption{An algorithm for computing $[\varphi_1 \TUntil{[a,b]} \varphi_2]_{\sigma}$}
  \label{algo:tuntil}
  \begin{algorithmic}
    \Require An FPC signal $[\varphi_2]_{\sigma}$ 
    given as a sequence $(t_{0},r_{0}),\dotsc,(t_{n},r_{n})$
    and a signal $g := [\BoxOp{[0,a]} \varphi_1 \UntilOp{} \varphi_2]_{\sigma}$ 
    as $(u_{0},v_{0}),\dotsc,(u_{m},v_{m})$
    \Ensure The FPL signal
    $[\varphi_1 \TUntil{[a,b]} \varphi_2]_{\sigma}$
    \State $t_{\mathsf{temp}} := \Max \{t_n - a, u_m \}$;
    \State $r_{\mathsf{left}} := [\varphi_2]_{\sigma}(t_{\mathsf{temp}} + a) \sqcap g(t_{\mathsf{temp}})$;
    \State $F := \bigl[ \;(t_{\mathsf{temp}}+a, r_{\mathsf{left}}) \;\bigr]$; 
    \Comment $F$ is the FPC signal $\tau \mapsto \Robust{\sigma^t}{\varphi_1 \UntilOp{[a,\tau]} \varphi_2}$
    \State $s := (b-a)r_{\mathsf{left}}$;
    \Comment The area of $F$
    \State $G := \bigl[ \;(t_{\mathsf{temp}}, s / (b-a), 0) \;\bigr]$; 
    \Comment $G$ is the FPC signal $[\varphi_1 \TUntil{[a,b]} \varphi_2]_{\sigma}$
    \While{$t_{\mathsf{temp}} \geq 0$}
    \State $t_{\mathsf{old}} := t_{\mathsf{temp}}$;
    \State $t_{\mathsf{temp}} :=$ the greatest $t$ such that 
    \State \qquad \qquad 
    $t < t_{\mathsf{old}} \wedge 
    \bigl(
    \exists t_i.\, t + a = t_i 
    \vee \exists (t', r') \in F. \; t+ b = t' 
    \vee \exists u_i.\, t = u_i \bigr)$;
    \State $\mathsf{Deq} := \{(t, r) \in F \mid t > t_{\mathsf{temp}} + b\}$; 
    \Comment{Dequeue old elements in $F$}
    \State $F := F \setminus \mathsf{Deq}$;
    \State $\mathsf{Pop} := \{(t, r) \in F \mid r \leq [\varphi_2]_{\sigma}(t_{\mathsf{temp}}+a) \}$; 
    \Comment{Pop small elements in $F$} 
    \State $F := F \setminus \mathsf{Pop}$;
    \State $t_{\mathsf{Pop}} := \mathsf{min} \{t \mid(t,r) \in F \; \text{ or } \; t = t_{\mathsf{temp}} + b\}$;
    \State $F := \big[ (t_{\mathsf{temp}}+a, r_{\mathsf{left}})\big] \cup F$;
    \Comment{Push the left endpoint of the window to $F$}
    \State $r_{\mathsf{Deq}} := \Max \{ r \mid (t,r) \in F\}$;
    \State $r_{\mathsf{Push}} := \Min \{ r \mid (t,r) \in F\}$;
    \State $s := s
    - (t_{\mathsf{old}} - t_{\mathsf{temp}})r_{\mathsf{Deq}} 
    - \mathsf{area}(\mathsf{Pop}) 
    + (t_{\mathsf{Pop}} - t_{\mathsf{temp}})r_{\mathsf{Push}}$;

    \State $\mathsf{Trunc} := \{(t, r) \in F \mid r > g(t_{\mathsf{temp}})\}$; 
    \Comment{Truncate big elements in $F$}
    \State $F := F \setminus \mathsf{Trunc}$;
    \State $t_{\mathsf{Trunc}} := \mathsf{min} \{t \mid(t,r) \in \mathsf{Trunc} \; \text{ or } \; t = t_{\mathsf{temp}} + b\}$;
    \State $F := F \cup \big[ (t_{\mathsf{Trunc}}, g(t_{\mathsf{temp}}))\big]$;
    \State $r_{\mathsf{left}} := \Min \{ r \mid (t,r) \in F \}$;
    \State $r_{\mathsf{right}} := \Max \{ r \mid (t,r) \in F \}$;
    \State $s := s
    - \mathsf{area}(\mathsf{Trunc}) 
    + ((t_{\mathsf{temp}} + b) - t_{\mathsf{Trunc}})r_{\mathsf{right}}$;
    \State $G := \{(t_{\mathsf{temp}}, s/(b-a),r_{\mathsf{right}} - r_{\mathsf{left}})\} \cup G$
    \EndWhile
  \end{algorithmic}
\end{algorithm}


\section{Omitted Proofs}
\subsection{Proof of Lem.~\ref{lemma:untilIsMonotone}}
\begin{myproof}
  We only prove the first inequality; the others are similar.
  \[
    \begin{array}{ll}
      \Robust{\sigma}{\varphi_1 \UntilOp{[t_0,t']} \varphi_2} 
      &= \Vee{\tau \in [t_0, t']}(\Robust{\sigma^{\tau}}{\varphi_1} \sqcap \Wedge{\tau' \in [0, \tau]} \Robust{\sigma^{\tau'}}{\varphi_2})\\
        & \geq \Vee{\tau \in [t_0, t]}(\Robust{\sigma^{\tau}}{\varphi_1} \sqcap \Wedge{\tau' \in [0, \tau]} \Robust{\sigma^{\tau'}}{\varphi_2})\\
            & = \Robust{\sigma}{\varphi_1 \UntilOp{[t_0,t]} \varphi_2} 
    \end{array}
  \]
  \myqed
\end{myproof}

\subsection{Proof of Lem.~\ref{lem:CorrespondenceBetweenUntilAndTUntil}}
\begin{myproof}
  We only show the proof of the first equality; the others are similar.
  
  We  first show (LHS) $\geq$ (RHS).
  \[
  \begin{array}{rlll}
    \text{(RHS)} 
    & = 
    & \Lim{t \to \infty} 
      \Frac{1}{t-t_0} \Int_{t_0}^{t} 
      \Robust{\sigma}{\varphi_1 \UntilOp{[t_0, \tau]} \varphi_2}^{+} d\tau 
    & \\
    & \leq 
    & \Lim{t \to \infty} \Frac{1}{t-t_0} \Int_{t_0}^{t}
      \Robust{\sigma}{\varphi_1 \UntilOp{[t_0, t]} \varphi_2}^{+} d\tau \qquad
    & \text{by Lem.~\ref{lemma:untilIsMonotone}}\\
    & = 
    & \Lim{t \to \infty}
      \Robust{\sigma}{\varphi_1\UntilOp{[t_0, t]}\varphi_2}^{+}
    &\\
    & \leq 
    & \Robust{\sigma}{\varphi_1\UntilOp{[t_0, \infty]}\varphi_2}^{+}&\\
    & = 
    & \text{(LHS)} 
    & \\ 
  \end{array}
  \]
  Now we show the equality (LHS) $=$ (RHS).
  Let 
  \[
    f(t) = 
    \Frac{1}{t-t_0} \Int_{t_0}^{t} 
    \Robust{\sigma}{\varphi_1 \UntilOp{[t_0, \tau]} \varphi_2}^{+} d\tau \;.
  \]
  By Lem.~\ref{lemma:untilIsMonotone}
  $\Robust{\sigma}{\varphi_1 \UntilOp{[t_0, \tau]} \varphi_2}^{+}$
  is monotonically increasing with respect to $\tau$,
  hence $f(t)$ is also monotonically increasing with respect to $t$
  because $f(t)$ is an average of $\Robust{\sigma}{\varphi_1 \UntilOp{[t_0, \tau]} \varphi_2}^{+}$
  over $t_0 \leq \tau \leq t$.
  If $f(t)$ is not bounded, then obviously (LHS) $\leq$ (RHS).
  Otherwise, if $f(t)$ is bounded,
  the increasing function $f(t)$
  converges to some $\alpha \in \Rnn$ as $t\to\infty$.
  By (LHS) $\geq$ (RHS) (that we have already shown),
  $\alpha + \varepsilon = $ (LHS) for some $\varepsilon \in \Rnn$.
  Here the following statement holds.
  \begin{equation}\label{eq:boundOfRHS}
    \forall \alpha' < \alpha + \varepsilon. \;
    \exists t' \in [t_0, \infty). \;
    \Robust{\sigma^{t'}}{\varphi_2} \sqcap
    \Wedge{\tau \in [0, t']} \Robust{\sigma^{\tau}}{\varphi_1} > \alpha'
  \end{equation}
  Hence, for such $\alpha'$ and $t'$,
  \[
    \begin{array}{rlll}
      \text{(RHS)} 
      & = & 
      \Lim{t \to \infty} \Frac{1}{t-t_0} \Int_{t_0}^{t} \Robust{\sigma}{\varphi_1 \UntilOp{[t_0, \tau]} \varphi_2}^{+} d\tau & \\
      & = & 
      \Lim{t \to \infty} \Frac{1}{t-t_0} \Big( 
      \Int_{t_0}^{t'} \Robust{\sigma}{\varphi_1 \UntilOp{[t_0, \tau]} \varphi_2}^{+} d\tau 
      + \Int_{t'}^{t} \Robust{\sigma}{\varphi_1 \UntilOp{[t_0, \tau]} \varphi_2}^{+} d\tau \Big)& \\
      & \geq & 
      \Lim{t \to \infty} \Frac{1}{t-t_0} 
      \Int_{t'}^{t} \Robust{\sigma}{\varphi_1 \UntilOp{[t_0, \tau]} \varphi_2}^{+} d\tau & \\
      & \geq & 
      \Lim{t \to \infty} \Frac{t-t'}{t-t_0} \alpha'  \qquad\text{by~(\ref{eq:boundOfRHS})}\\
      & = & \alpha'\enspace. & \\
    \end{array}
  \]
  Therefore we have
  \[
    \forall \alpha' < \alpha + \varepsilon. \qquad \text{(RHS)} = \alpha \wedge \text{(RHS)} \geq \alpha'    
  \]
  and hence $\varepsilon = 0$.
  Consequently
  (LHS) $=$ (RHS).
  \myqed
\end{myproof}

\subsection{Proof of Prop.~\ref{prop:diaRefinement} and~\ref{prop:BoxRefinement}}
 We start with the following lemmas.
\begin{mylemma}[logical monotonicity]
  \label{lem:monotonicity}
  Let $\C$ be a positive context (Def.~\ref{def:context}). We have
  \begin{align*}
    \begin{array}{rcl}
      \forall \sigma.\; \Robust{\sigma}{\varphi}^{+} 
      \leq \Robust{\sigma}{\varphi'}^{+}
      \quad&\text{implies}
      &\quad 
        \forall \sigma.\; \Robust{\sigma}{\C[\varphi]}^{+} 
        \leq
        \Robust{\sigma}{\C[\varphi']}^{+}\enspace;\;\text{and}\\
      \forall \sigma.\; \Robust{\sigma}{\varphi}^{-} 
      \leq \Robust{\sigma}{\varphi'}^{-}
      \quad&\text{implies}
      &\quad 
        \forall \sigma.\; \Robust{\sigma}{\C[\varphi]}^{-} 
        \leq 
        \Robust{\sigma}{\C[\varphi']}^{-} \enspace.
    \end{array}  
  \end{align*}
\end{mylemma}
\begin{myproof}
  By induction on the construction of the positive context $\C$.
  \myqed
\end{myproof}

\begin{mylemma}\label{lem:zeroCorrespondence}
  Let $\varphi, \varphi'$ be $\AvSTL$ formulas and $\C$ be a positive context. Then
  \[
    \forall \sigma. \Robust{\sigma}{\varphi}^{+} > 0 
    \implies
    \forall \sigma. \Robust{\sigma}{\varphi'}^{+} > 0 
  \]
  implies
  \[
    \forall \sigma. \Robust{\sigma}{\C[\varphi]}^{+} > 0 
    \implies
    \forall \sigma. \Robust{\sigma}{\C[\varphi']}^{+} > 0 
  \]
\end{mylemma}
\begin{myproof}
  Straightforward by induction on the construction of $\C$.
  \myqed
\end{myproof}

Now we prove Prop.~\ref{prop:diaRefinement}, soundness and completeness of $\Diamond$-refinement.
\begin{myproof} (Of Prop.~\ref{prop:diaRefinement})
  Obviously we have
 \begin{math}
    \Robust{\sigma}{\TDiaOp{[a,b]}\varphi}^{+} < \Robust{\sigma}{\DiaOp{[a,b]}\varphi}^{+}
  \end{math}; therefore by Lem.~\ref{lem:monotonicity}, we have  
  \begin{displaymath}
    \Robust{\sigma}{\C[\TDiaOp{[a,b]}\varphi]}^{+} > 0 
    \quad\implies\quad
    \Robust{\sigma}{\C[\DiaOp{[a,b]}\varphi]}^{+} > 0\enspace.
  \end{displaymath}
  To prove the opposite direction, by Lem.~\ref{lem:zeroCorrespondence}, 
  it suffices to show the following.
  \[
  \Robust{\sigma}{\DiaOp{[a,b']}\varphi}^{+} > 0
  \implies
  \Robust{\sigma}{\TDiaOp{[a,b]}\varphi}^{+} > 0 
  \quad \text{for any $b' < b$}\; .
  \]
  Assume $\Robust{\sigma}{\DiaOp{[a,b']}\varphi}^{+} > 0$.
  Then
  \begin{align*}
    &\Robust{\sigma}{\TDiaOp{[a,b]} \varphi}^{+}\\
    & = \Frac{1}{b-a} \Int_{a}^{b} \Robust{\sigma}{\DiaOp{[a,\tau]} \varphi}^{+} \;d\tau
    & \text{ by the definition of $\TDiaOp{}$}\\
    & = \Frac{1}{b-a} 
      \bigg( \Int_{a}^{b'} \Robust{\sigma}{\DiaOp{[a,\tau]} \varphi}^{+} \;d\tau
      + \Int_{b'}^{b} \Robust{\sigma}{\DiaOp{[a,\tau]} \varphi}^{+} \;d\tau \bigg)
    & \\
    & \geq \Frac{1}{b-a} 
      \bigg( \Int_{a}^{b'} 0 \; d\tau
      + \Int_{b'}^{b} \Robust{\sigma}{\DiaOp{[a,b']} \varphi}^{+} \;d\tau \bigg)
    & \text{ by Prop.~\ref{lemma:untilIsMonotone} and $b' < b$}\\
    & = \Frac{b-b'}{b-a} \Robust{\sigma}{\DiaOp{[a,b']}\varphi}^{+} > 0 \;. \tag*{\myqed}\\
  \end{align*}
\end{myproof}

Then we prove Prop.~\ref{prop:BoxRefinement}, soundness and completeness
of $\Box$-refinement.
\begin{myproof} (Of Prop.~\ref{prop:BoxRefinement})
  From Lem.~\ref{lem:monotonicity}, 
  \begin{displaymath}
    \Robust{\sigma}{\C[\BoxOp{[a,b]}\varphi \wedge \TBoxOp{[b,b+\delta]}\varphi]}^{+} > 0 
    \quad\implies\quad
    \Robust{\sigma}{\C[\BoxOp{[a,b]}\varphi]}^{+} > 0
  \end{displaymath}
  is obvious.
  We want to show the other direction.
  From Lem.~\ref{lem:zeroCorrespondence},
  it suffices to show 
  \begin{displaymath}
    \Robust{\sigma}{\BoxOp{[a,b']}\varphi}^{+} > 0 
    \quad\implies\quad
    \Robust{\sigma}{\BoxOp{[a,b]}\varphi \wedge \TBoxOp{[b,b+\delta]}\varphi}^{+} > 0
  \end{displaymath}
  for any $b' > b$.
  Here $\BoxOp{[a,b']} \varphi \cong \BoxOp{[a,b]} \varphi \wedge \BoxOp{[b,b']} \varphi$,
  hence 
  the above implication holds if so does the following.
  \begin{displaymath}
    \Robust{\sigma}{\BoxOp{[b,b']}\varphi}^{+} > 0 
    \quad\implies\quad
    \Robust{\sigma}{\TBoxOp{[b,b+\delta]}\varphi}^{+} > 0 \;.
  \end{displaymath}  
  In the case of $b' \geq b+ \delta$,
  it obviously holds.
  Otherwise,
  in the case of $b < b' < b+ \delta$,
  we proceed as follows.
  Assume $\Robust{\sigma}{\BoxOp{[b,b']}\varphi}^{+} > 0$.
  Then 
  \begin{align*}
    &\Robust{\sigma}{\TBoxOp{[b,b+\delta]} \varphi}^{+}\\
    & = \Frac{1}{\delta} \Int_{b}^{b+\delta} \Robust{\sigma}{\BoxOp{[b,\tau]} \varphi}^{+} \;d\tau
    & \text{ by the definition of $\TBoxOp{}$}\\
    & = \Frac{1}{\delta} 
      \bigg( \Int_{b}^{b'} \Robust{\sigma}{\BoxOp{[b,\tau]} \varphi}^{+} \;d\tau
      + \Int_{b'}^{b+\delta} \Robust{\sigma}{\BoxOp{[b,\tau]} \varphi}^{+} \;d\tau \bigg)
    & \\
    & \geq \Frac{1}{\delta} 
      \bigg( \Int_{b}^{b'} \Robust{\sigma}{\BoxOp{[b,b']} \varphi}^{+} \;d\tau
      + \Int_{b'}^{b+\delta} 0 \;d\tau \bigg)
    & \text{ by Prop.~\ref{lemma:untilIsMonotone} and $b < b' < b + \delta$}\\
    & = \Frac{b' - b}{\delta} \Robust{\sigma}{\BoxOp{[b,b']}\varphi}^{+} > 0 \;. \tag*{\myqed}\\
  \end{align*}
\end{myproof}

\subsection{Proof of Thm.~\ref{thm:complexity_rel}}
\begin{myproof} 
 We obtain the robustness value
 $\Robust{\sigma}{\varphi}^{+}$
 via the robustness signals $[\psi]^{+}_{\sigma}$ for subformulas $\psi$
 of $\varphi$. This is done by induction on $\psi$.
 
 Before we hit an averaged modality we use the algorithm from~\cite{DBLP:conf/cav/DonzeFM13} (described
 in~\S{}\ref{subsec:algoSTL}).
 Note that all the signals that we deal with are
 finitely piecewise-constant; by
 analyzing~\cite[Thm.~3]{DBLP:conf/cav/DonzeFM13},
 it is easy to see that the computation of
$[\psi]^{+}_{\sigma}$ has time-complexity in
 $\mathcal{O}(|\psi| |\sigma|)$. Furthermore, the size of 
 $[\psi]^{+}_{\sigma}$ (in the sense of Def.~\ref{def:signal}) is 
 in $\mathcal{O}(|\sigma|)$.

 Once we hit an averaged modality (like $\TDiaOp{I}$ or $\TUntil{I}$),
 it is
 taken care of by  Algorithm~\ref{algo:tdia} (for $\TDiaOp{I}$), 
 Algorithm~\ref{algo:tuntil} (for $\TUntil{I}$) and their adaptations
 (for $\TBoxOp{I}$ and $\TRelease{I}$). The time-complexity of the computation is
  $\mathcal{O}(|\psi| |\sigma|)$, and the resulting 
 signal $[\psi]^{+}_{\sigma}$ has the size 
 in $\mathcal{O}(|\sigma|)$.
The difference, however, is that the robustness signal
 $[\psi]^{+}_{\sigma}$ is no longer finitely piecewise-constant but is
 piecewise-linear.

 After that we again apply the algorithm from~\cite{DBLP:conf/cav/DonzeFM13} (see~\S{}\ref{subsec:algoSTL}),
 but now to the input signal that is finitely
 piecewise-\emph{linear}. In this case, the
 time-complexity as well as the size of  $[\psi]^{+}_{\sigma}$ is shown to be
 in $\mathcal{O}(d^{|\psi|}  |\psi| |\sigma|)$~\cite[Thm.~3]{DBLP:conf/cav/DonzeFM13}. The extra factor
$d^{|\psi|}$ is due to the extra timestamped values that arise from 
  two sloped lines crossing each other.
\auxproof{
  First, in the case of $\varphi$ is averaging-free,
  then the statement is obviously true from Thm.~\ref{thm:complexity}.
  Then, let us consider 
  the case of $\varphi \equiv \TDiaOp{I}{\varphi'}$ 
  where $\varphi'$ is averaging-free.
  Because $\varphi'$ is averaging-free,
  we can calculate $[\varphi']_\sigma$ 
  with time-complexity in 
  $\mathcal{O}(d^{|\varphi'|} |\varphi'| |\sigma|)$
  from Thm.~\ref{thm:complexity}.
  Moreover,
  the number of the timestamps of $[\varphi']_\sigma$
  is bounded above by $d^{|\varphi'|} |\sigma|$.
  See Theorem~3 in \cite{DBLP:conf/cav/DonzeFM13}.
  Hence the function
  $[\TDiaOp{I}{\varphi}]_\sigma$
  is computed in time $\mathcal{O}(d^{|\varphi'|} |\sigma|)$
  from Prop.~\ref{prop:complexity_tdia}.
  By the construction,
  the number of the timestamps of $[\varphi]_\sigma$
  is bounded above by $2(d^{|\varphi'|} |\sigma|)$.
  Now we apply Thm.~\ref{thm:complexity},
  and obtain the function $[\C[\varphi]]_\sigma$.
  The time-complexity of this step is
  \begin{equation}
    \mathcal{O}(d^{|\C|} |\C| |[\varphi]_\sigma|) 
    = \mathcal{O}(d^{|\C|} |\C| d^{|\varphi'|} |\sigma|) 
    = \mathcal{O}(d^{|\C|+|\varphi'|} |\C| |\sigma|).
  \end{equation}
  Hence the total time-complexity is 
  \begin{align*}
    &\mathcal{O}(d^{|\varphi'|} |\varphi'| |\sigma| + d^{|\varphi'|} |\sigma| + d^{|\C|+|\varphi'|} |\C| |\sigma|)\\
    \subseteq &\mathcal{O}(d^{|\C|+|\varphi'|} (|\varphi'| |\sigma| + 1 + |\C| |\sigma|)\\
    \subseteq &\mathcal{O}(d^{|\C|+|\varphi'|} (|\varphi'| + 1 + |\C|) |\sigma|)\\
    \subseteq
 &\mathcal{O}(d^{|\C[\varphi]|} |\C[\varphi]| |\sigma|).\\
  \end{align*}
  Finally, about the case of $\varphi \equiv \TBoxOp{I}{\varphi'}$,
  we can proof in the same manner.
}  \myqed
\end{myproof}

\section{ The Automatic Transmission
 Model~\cite{HoxhaAF14arch1}}\label{appendix:ATModel}
\begin{figure}[hbp]
   \begin{minipage}{.48\textwidth}
     \includegraphics[width=\textwidth]{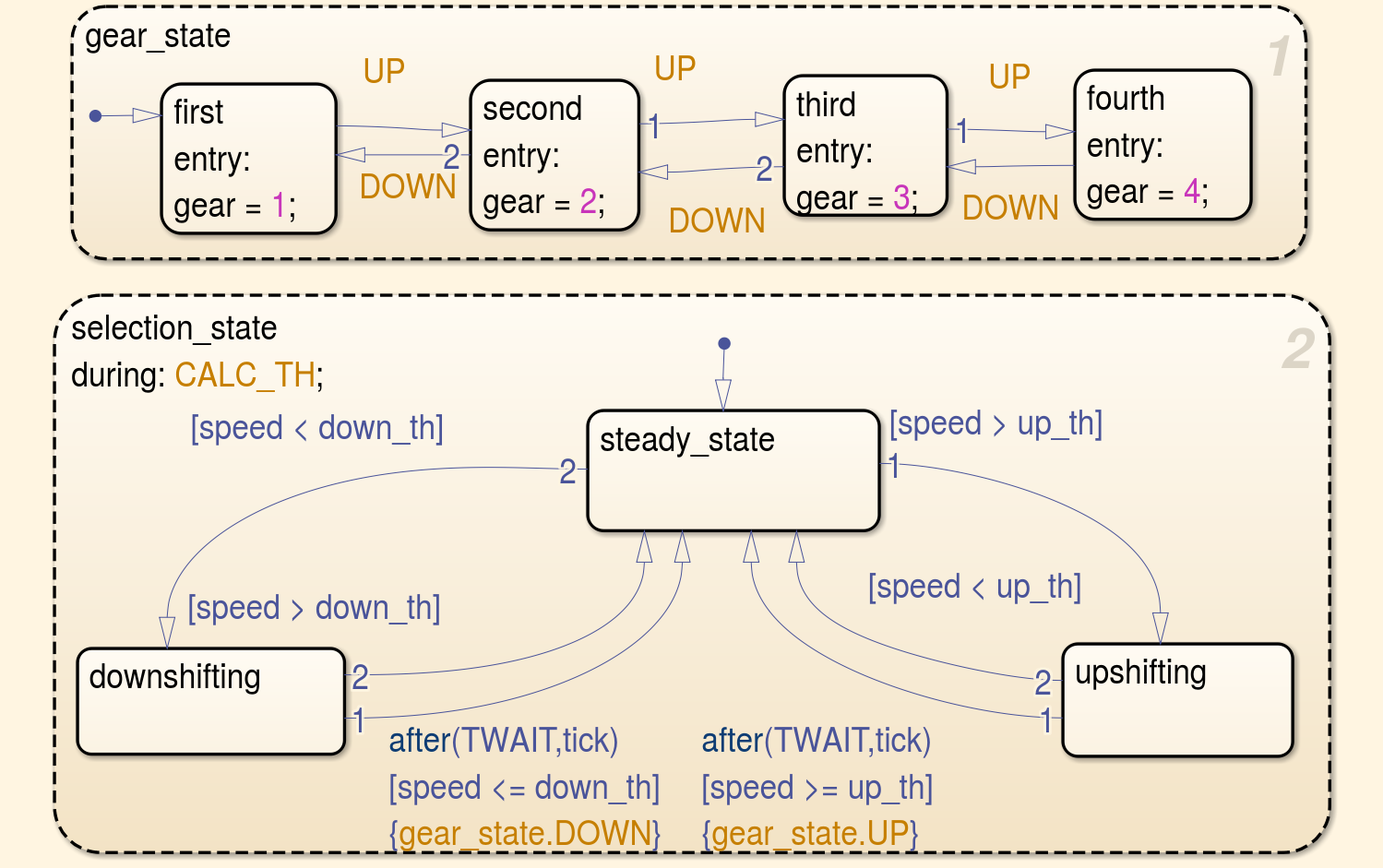}
     \caption{The automatic transmission model
       from~\cite{HoxhaAF14arch1}: the Stateflow diagram for the digital
       controller of the gear}
     \label{fig:shiftlogic}
   \end{minipage}\quad
   \begin{minipage}{.45\textwidth}
     \includegraphics[trim=0cm 0cm 0cm 2.2cm, width=\textwidth]{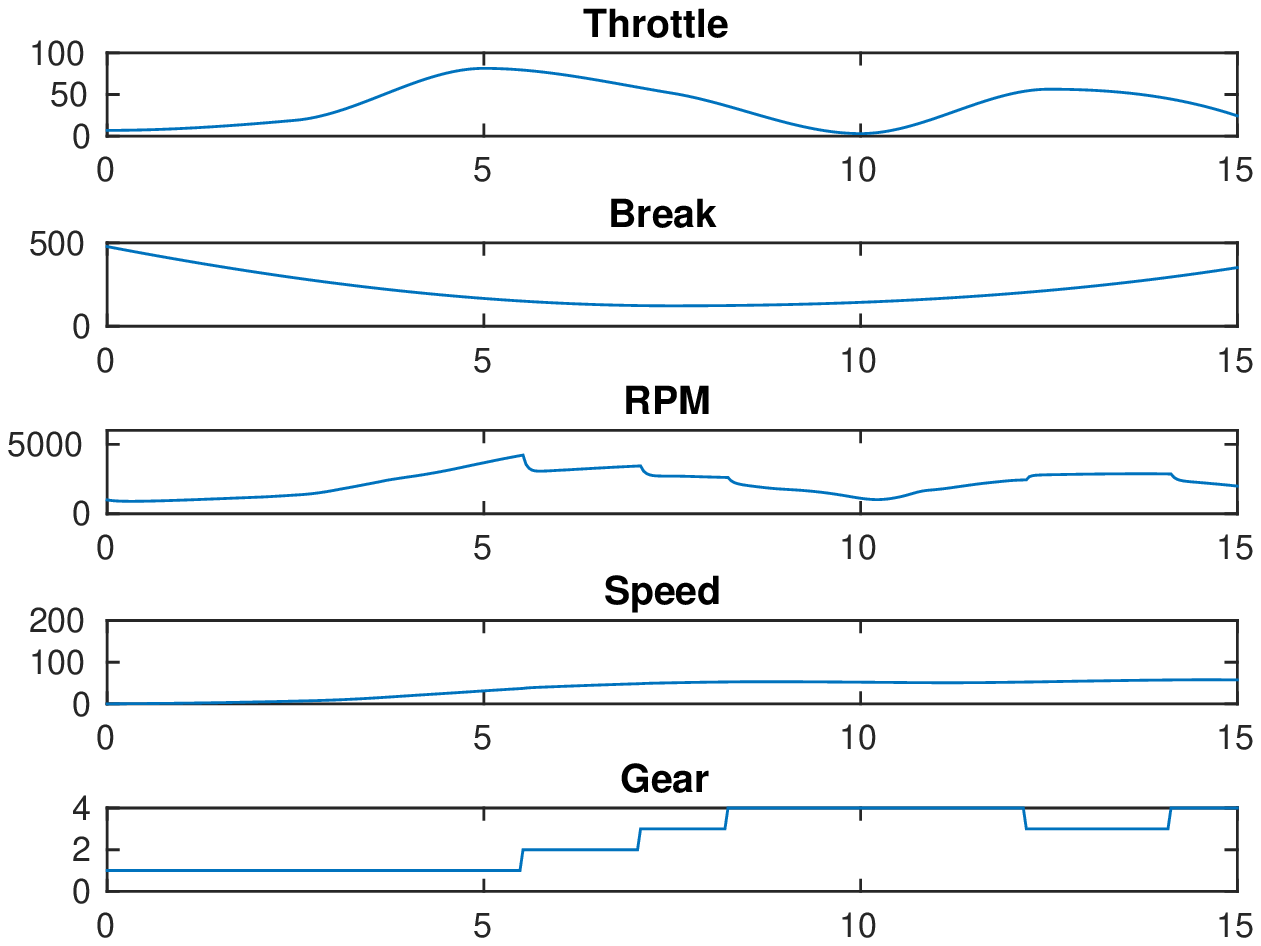}
     \caption{The automatic transmission model
       from~\cite{HoxhaAF14arch1}: a   trajectory example}
     \label{fig:trajectory}
   \end{minipage}
   \centering
   \includegraphics[trim=0cm 2cm 0cm 0cm, width=.7\textwidth]{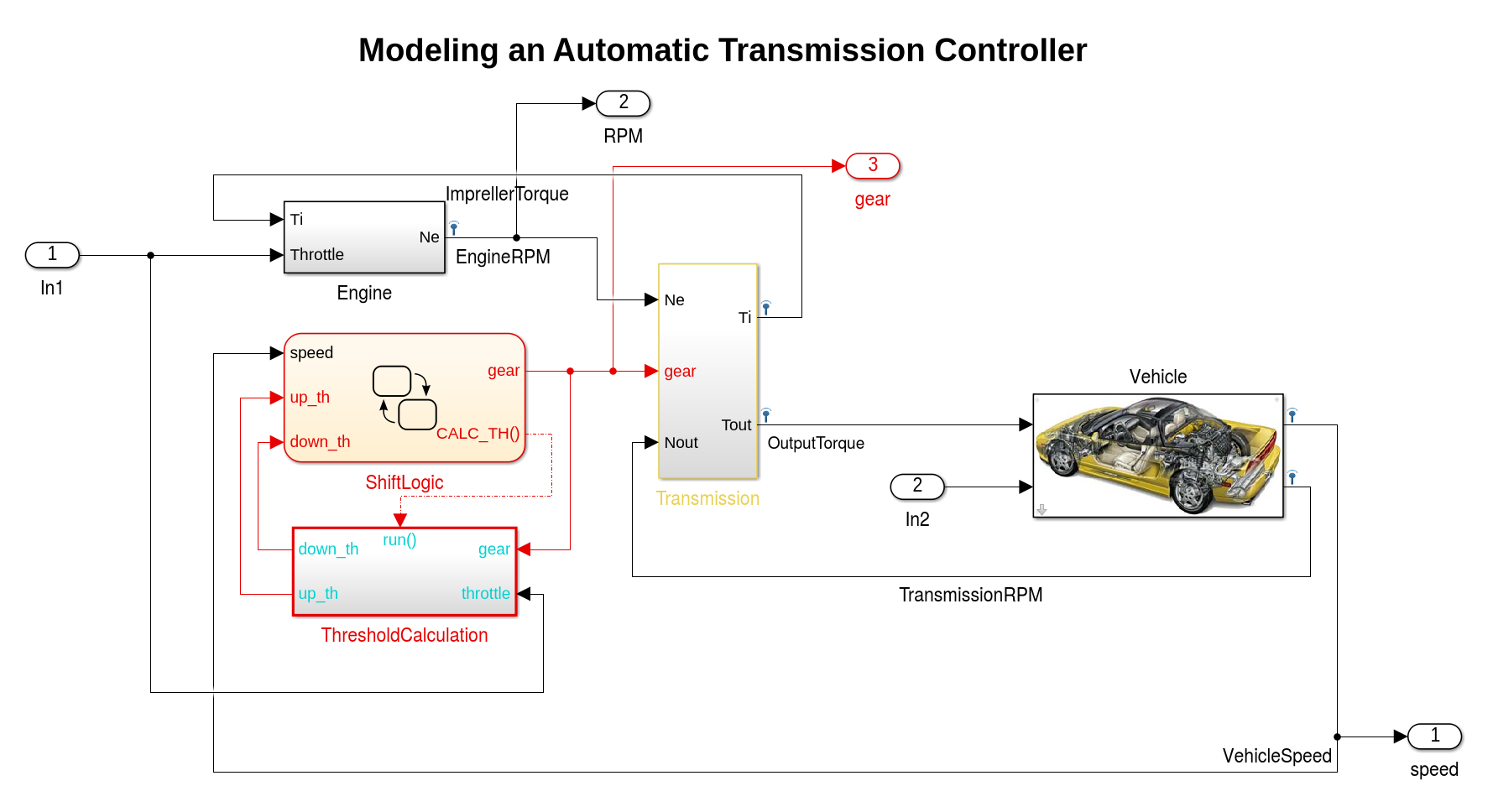}
   \caption{The automatic transmission model
     from~\cite{HoxhaAF14arch1}: the Simulink diagram}\label{fig:AT}
\end{figure}

The model is given by a Simulink diagram in Fig.~\ref{fig:AT}; therein
the block for the digital
   controller of the gear is realized as a Stateflow diagram in
   Fig.~\ref{fig:shiftlogic}. An example of the model's trajectories is
   in Fig.~\ref{fig:trajectory}.

\end{document}